\documentclass[10pt,reqno,final]{amsart}
\pdfoutput=1
\usepackage{}
\usepackage{amsfonts}
\usepackage{amssymb}
\usepackage{epsfig,amssymb,amsmath,version}
\usepackage{amssymb,version,graphicx,fancybox,mathrsfs,stmaryrd,url}
\usepackage[notcite,notref]{showkeys}
\usepackage{subfigure}
\usepackage{color}
\usepackage{cases}
\usepackage{makecell}
\usepackage{multirow}

\newcommand{\bs}[1]{\boldsymbol{#1}}
\def \ri {{\rm i}}
\def \ce {{\rm ce}}
\def \se {{\rm se}}

\def \Sum {\sum_{|m|=0}^{\infty}}

\def \Hankel {H_m ^{(1)}}

\begin{document}


\bibliographystyle{plain}
\baselineskip 13pt

{\title[Circular and elliptic cylindircal  cloaks] {Accurate  Simulation of Ideal   Circular and Elliptic Cylindrical  Invisibility  Cloaks}
\author[Z. Yang\;\;  $\&$\;\;  L. Wang] {Zhiguo Yang \;
and\; Li-Lian Wang}
\thanks{\noindent Division of Mathematical Sciences, School of Physical
and Mathematical Sciences,  Nanyang Technological University,
637371, Singapore. The research of the authors is  supported by a Singapore  MOE AcRF Tier 2 Grant (MOE 2013-T2-1-095, ARC 44/13),  a Singapore A$^\ast$STAR-SERC-PSF Grant (122-PSF-007) and a Singapore MOE AcRF Tier 1 Grant (RG 15/12).
}

\begin{abstract}  The coordinate transformation offers a remarkable way to design cloaks that  can steer electromagnetic fields so as to  prevent waves from penetrating into
the {\em cloaked region} (denoted by $\Omega_0$, where the objects inside are invisible to observers outside).  The ideal circular and elliptic cylindrical cloaked regions  are blown up from a point and a line segment, respectively,  so  the transformed material parameters and the corresponding coefficients of the resulted equations are highly singular at the cloaking boundary  $\partial \Omega_0$.   The electric field or magnetic field is not continuous across $\partial\Omega_0.$   The imposition of appropriate {\em cloaking boundary conditions} (CBCs)   to achieve perfect concealment  is a crucial but  challenging issue.

Based upon the principle that  finite  electromagnetic fields in the original space must be  finite  in the transformed space as well,  we obtain  CBCs   that intrinsically relate to the essential ``pole'' conditions  of a singular transformation.   We also find that for the elliptic  cylindrical cloak,  the CBCs should be imposed differently for the cosine-elliptic  and sine-elliptic  components of  the decomposed fields.  With these at our disposal,  we can rigorously  show that the governing equation in $\Omega_0$ can be decoupled from the exterior region $\Omega_0^c$, and the total  fields in the cloaked region vanish.  We emphasize that our proposal of CBCs is different from any existing ones.

Using  the exact circular (resp.,  elliptic)  Dirichlet-to-Neumann (DtN) non-reflecting boundary conditions
to  reduce the unbounded  domain $\Omega_0^c$ to a bounded domain,
we introduce  an accurate  and efficient Fourier-Legendre spectral-element method (FLSEM)   (resp.,  Mathieu-Legendre spectral-element method (MLSEM)) to simulate  the circular cylindrical cloak (resp.,  elliptic cylindrical cloak).          We provide ample numerical  results to demonstrate that  the perfect concealment of waves  can be achieved for the ideal circular/elliptic cylindrical cloaks under our proposed  CBCs and accurate numerical solvers.
\end{abstract}
 \maketitle

\vspace*{-12pt}

\section{Introduction}

Since the groundbreaking  works of  Pendry, Schurig and Smith  \cite{Pendry.2006},   and Leonhardt
\cite{leonhardt2006optical},  transformation  optics (or transformation electromagnetics)
has emerged as an unprecedentedly powerful tool  for metamaterial design (see
\cite{cui2009metamaterials,cai2010optical,werner2013tran} and  many original references therein).
The use of coordinate transformations has also been  explored earlier by
Greenleaf et al.  \cite{greenleaf2003nonuniqueness} in the context of electrical impedance tomography.
Perhaps, one of the most appealing applications of metamaterials is the invisibility cloak \cite{leonhardt2012geometry}.  The  mechanism of   a cloak  is  typically  based on a  singular  coordinate transformation of the Maxwell  equations that can steer
the electromagnetic waves without penetrating  into  the cloaked region, and  thereby render  the interior effectively ``invisible'' to the outside
\cite{Pendry.2006}.   The first experimental demonstration of
a two-dimensional  cloak with a simplified model  was realized by Schurig et al.   \cite{schurig2006metamaterial}, along with
 full-wave finite-element simulations  
 \cite{cummer2006full,zolla2007electromagnetic}.  These impactive works have inspired a surge of
 developments and innovations (see  \cite{zhang2012electrodynamics,FleAlu14,Kwon14}  for an up-to-date review).

In this paper, we are largely concerned with  mathematical and numerical study of the  ideal circular cylindrical  cloak using the transformation in Pendry et al.
   \cite{Pendry.2006} and its important variant, i.e., the elliptic cylindrical cloak  \cite{ma2008material,cojocaru2009exact}.
   The coordinate transformation in  \cite{Pendry.2006} suppresses a disk into an annulus so that
  the interior ``empty'' space,  constituting the  cloaked region (see Figure \ref{cloakA}).   Such a   ``point-to-circle'' blowup
  leads to the    electric permittivity and
    magnetic permeability singular  at the  {\em inner boundary}  (denoted by $r=R_1$) of the cloak.   Accordingly,  the coefficients of the  governing equation are highly singular.
   The presence of singularities  poses significant  challenges for   simulation,  realization and analysis as well.   A critical issue resides in  {\em how to impose suitable conditions at the inner boundary, i.e., CBCs,  to achieve perfect concealment of waves.}
   We highlight below some relevant studies and attempts, which are  by no means comprehensive, given  a large volume of existing literature.
   \begin{itemize}
    \item Ruan et al. \cite{ruan2007ideal}  first analytically studied the sensitivity of the ideal cloak  \cite{Pendry.2006} to a small $\delta$-perturbation of the inner  boundary  (i.e.,  from $R_1$ to $R_1+\delta,$ while the material parameters  remained unchanged) under  the transverse-electric (TE) polarization.
    Their findings are  (i) the ideal cloak in  \cite{Pendry.2006}  is sensitive to a tiny perturbation  of the boundary;
     (ii) the electric field is discontinuous across the inner boundary; and
     (iii) the perturbed cloak is nearly ideal
      in the sense that  the magnitude of the  fields penetrated into the cloaked region is  small.
  \item  Zhang et al. \cite{zhang2007response} provided    deep  insights into the physical effects, and found that the  singular transformation gave rise to electromagnetic surface currents along the inner interface of the ideal cloak (also see \cite{zhang2012electrodynamics}).     

\item To shield the incoming  waves,  the perfect magnetic conductor  {\rm(PMC)}  condition (i.e., the tangential component of the magnetic field vanishes) was  imposed at $r=R_1$ in finite-element simulations (see,  e.g.,  \cite{cummer2006full,Jichun.2012,ma2008material}). Indeed, such a condition can be naturally  implemented by using  N{\'e}d{\'e}lec edge-elements \cite{Nedelec80} in the Cartesian coordinates (see,  e.g., \cite{Monk03,li2012time}).  However, in the polar coordinates,  the PMC condition is  automatically satisfied, so it  does not lead to an  independent condition (see Remark $2.2$).

\item Weder \cite{Weder08} proposed CBCs for general point transformed cloaks from the perspective of  energy conservation.  Its implication to the three-dimensional ideal  spherical cloak  by  Pendry  et al.  \cite{Pendry.2006} is  that  the tangential components of the electric and  magnetic fields have to vanish at  the spherical surface $\partial\Omega^+_0,$
and that  the normal components of the curl of both fields have to vanish at the inner spherical surface $\partial\Omega^-_0.$ Under this set of CBCs,
the interior fields (i.e., in $\Omega_0$)  are  decoupled from  the exterior fields.  However,  CBCs in \cite{Weder08} are  not  applicable to the  ideal  circular cloak, as  the tangent component of the electric field does not vanish at $r=R_1^+$ (cf. \cite{ruan2007ideal,zhang2007response}).

 \item Lassas  and Zhou \cite{LaZhou11,LaZhou14}  proposed some non-local pesudo-differential CBCs  from a limiting process of non-singular approximate   two-dimensional Helmholtz cloaking. Their findings also indicate that   CBCs for two dimensions and three dimensions take  different forms,  and their  physical effects on the cloak interface are very different as well.
 \end{itemize}

In this paper, we propose CBCs based on the principle that   {\em the finite electric and magnetic fields in the original coordinates must be finite near the inner boundary within the cloak, after transformation}.    This situation is reminiscent to  the  imposition of   ``pole'' conditions  associated with the polar transformation (see, e.g., \cite{Got.O77, Shen97,Boyd01}).  The polar transformation is  singular at the origin, so additional conditions should be imposed so as to have desired regularity when the solution is transformed back to  the
Cartesian coordinates.     The {\em essential ``pole'' conditions} are the sufficient and necessary  conditions for spectral accurate simulations,  in other words, ignoring these will lead to inaccurate results (cf.
\cite{Shen97,ShenTaoWang2011}).
This notion has been  extended to study  other singular transformations, e.g., the spherical transformation and Duffy transformation (cf. \cite{shen2009triangular}). 
With this principle at our disposal, we obtain the desired CBCs from  the essential ``pole'' conditions of the singular transformation \cite{Pendry.2006}  at $r=R_1^+, $ and the  continuity of tangential component of  the magnetic field.   We find that  for the circular cylindrical cloak,  the cloaked region is decoupled from the
exterior,  and the total field therein is zero.    This also admits  the ``{\em finite energy}'' solution in some weighted Sobolev space in the transformed coordinates.

Compared with the circular case, the elliptic cloak is much less studied.  
The singular transformation \cite{ma2008material,cojocaru2009exact},  blows up  a line segment to an ellipse with foci being  the endpoints of the line segment.  As a result,
the ``line-to-ellipse''  transformation is only singular at two points, as opposite to the circular case.
Using the aforementioned principle for CBCs,
 we need to  decompose the full wave into the ``cosine-elliptic'' and ``sine-elliptic'' waves,  and the  essential ``pole'' conditions  must be  imposed differently for two components.

   This paper also aims at providing super-accurate numerical solvers  for simulating the  ideal circular and elliptic cylindrical  cloaks.  In full-wave finite-element simulations,  the perfect matched layer (PML) technique,   originated from  \cite{Bere94},  is  mostly used  to reduce the unbounded domain to a bounded one.
   For accurate  simulations (in particular, when the  frequency of the incident wave is high), we find it's beneficial, perhaps necessary,  to  employ the exact circular/elliptic DtN  non-reflecting boundary conditions (cf. \cite{Gro.K95}).   Indeed, the exact DtN boundary has been efficiently integrated with the
   spectral-Galerkin methods for wave scattering simulations  (see,  e.g.,  \cite{Shen.W05,FNS07,FSW09,Wang2Zhao12}).
   Benefited from the   separable geometry of the cloaks and the use of DtN boundary conditions,  we are able to
   employ Fourier/Mathieu expansions in angular direction and then use the Legendre spectral-element method to {numerically
   solve} the one-dimensional problems in radial direction.   We demonstrate that the proposed direct solver is fast,  accurate and robust for high-frequency   waves.  More importantly,  the produced numerical results show the perfectness of the ideal cloaks. 

It is important to remark that various interesting  approximate cloaks have been  proposed in    e.g.,
\cite{Greenleaf2009cloaking,Kohn10,Liu2011approximate,Liu2012enhanced,Liu2013enhanced,Ammari2013enhancement},  and that   the  time-domain simulations have been  attracting  much recent attention  (see, e.g., \cite{Hao2008fdtd,li2012time,Li2012developing,Li2014wellpose}).

%

 The  rest of this paper is organized as follows.    In section \ref{sect:circular}, we study the
 ideal circular cylindrical cloak. We start with formulating the governing equation including  CBCs and DtN non-reflecting boundary conditions, and then show that the  field  in the cloaked region vanishes. Finally, we  describe the   FLSEM for numerical simulations.      In section  \ref{sect:elliptic},   we focus on the mathematical and numerical study of   the elliptic cylindrical cloak.    In section \ref{sect:numer}, we provide numerous  simulation results to
 demonstrate  the perfectness of the ideal cloaks based on our proposed CBCs and numerical solvers.
 We conclude the paper with some remarks.

\section{Circular cylindrical cloaks}\label{sect:circular}
In this section, we  formulate the problem  that models the ideal circular cylindrical cloak
 and describe the FLSEM for its numerical simulation.  We put the emphasis  on the imposition of CBCs. 

\subsection{Coordinate transformation}
The cylindrical cloak   is based on  the coordinate transformation  in  Pendry et al.  \cite{Pendry.2006}, which   compresses
the cylindrical region  $\rho<R_2$   into the cylindrical  annular region $0<R_1<r<R_2,$ and takes the form
\begin{equation}\label{TransMap}
r=\frac{R_2-R_1}{R_2} \rho +R_1,\;\;\; \theta=\theta, \;\;\;  z=z,
\end{equation}
where $(\rho,\theta, z)$ is the cylindrical coordinates in the original space, and
$(r,\theta, z)$ is the cylindrical coordinates in the virtual  space (i.e., transformed space).    

\vspace*{-10pt}

\begin{figure}[!htb]
  \begin{minipage}[b]{0.49\textwidth}
 \baselineskip 16pt
\qquad The origin   is mapped to the circle $r=R_1$  that produces an ``empty'' space: $0\le r< R_1,$  forming the ``cloaked region'' to conceal any object inside.    The annulus $R_1<r<R_2$ constitutes the ``cloak'',  where the material parameters are obtained by applying the transformation \eqref{TransMap} to the Maxwell equations.    The exact DtN boundary condition is imposed at  $r=R_3$ to reduce the unbounded computational domain, and the material parameters in the outmost annulus  are positive constants (see  Figure \ref{cloakA}).
  \end{minipage} %
  \begin{minipage}[b]{0.5\textwidth}
    \centering
    \includegraphics[width=1\textwidth]{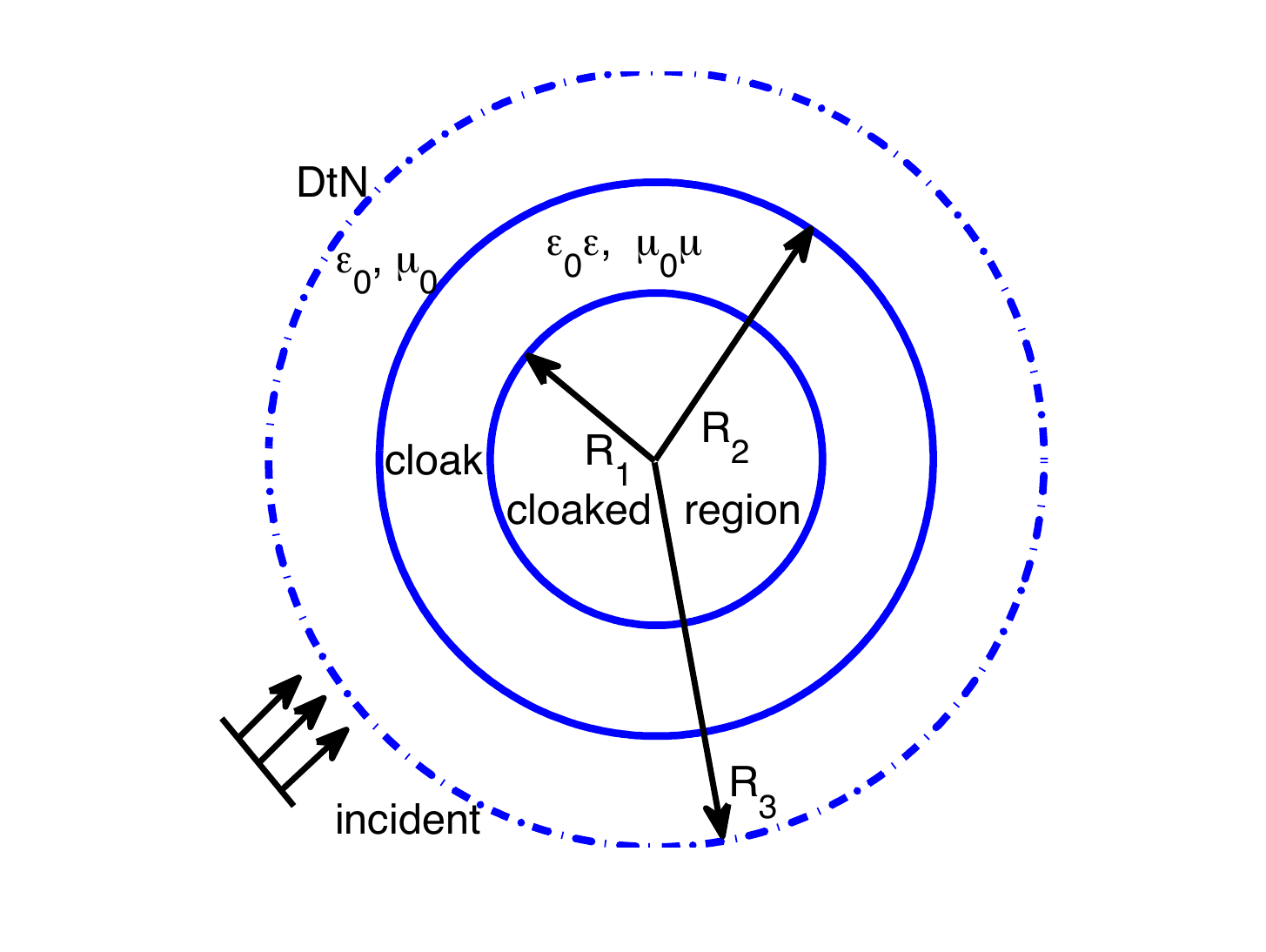}
    \caption{Schematic illustration of the circular cloak}
 \label{cloakA}
  \end{minipage} %
  \end{figure}


\vspace*{-18pt}

 Consider the free space time-harmonic Maxwell equations with   angular frequency $\omega$ in the original   $(\rho,\theta, z)$-coordinates:
\begin{equation} \label{time_harmonic_Max}
\nabla \times {\bs E}-\ri \omega \mu_0 {\bs H}=0,\quad \nabla \times {\bs H} + \ri \omega \varepsilon_0 {\bs E} =0,  \;\;\;  {\rm in} \;\; {\mathbb R}^3,
\end{equation}
where  a $e^{-{\rm i}\omega t}$ (note: $\ri$ is the complex unit) time dependence is assumed, and  the permeability $\mu_0$ and  permittivity $\varepsilon_0$ are  positive constants. It is known that  the Maxwell equations are form invariant under coordinate transformations.
  Following the approach in \cite{Pendry.2006},  we apply the mapping
\eqref{TransMap},  together with an identity transformation for $\rho>R_2,$  to  \eqref{time_harmonic_Max}, leading to the Maxwell  equation in the  new
$(r,\theta, z)$-coordinates:
\begin{equation} \label{mapped-Max}
\nabla' \times {\bs E}-\ri \omega \mu_0 {\bs \mu} {\bs H}=0,\quad \nabla' \times {\bs H} + \ri \omega \varepsilon_0 {\bs  \varepsilon} {\bs E} =0,   \quad r>R_1,
\end{equation}
with  anisotropic, singular material parameters  given by
\begin{align}
&{\bs  \varepsilon}= {\bs  \mu}={\rm diag}(\varepsilon_r,\varepsilon_\theta,\varepsilon_z),\quad   {\rm if}\;\;\;  R_1<r<R_2,  \label{media1}\\
& {\bs  \varepsilon}=\bs \mu={\bs I_3},\qquad\qquad\qquad \; {\rm if}\;\;\; r>R_2,\label{media2}
\end{align}
where $\bs I_3$ is the $3\times 3$ identity matrix, and
\begin{equation}\label{media}
\varepsilon_r=\mu_r= \frac{r-R_1}{r},\quad  \varepsilon_\theta=\mu_\theta=\frac{r}{r-R_1},  \quad \varepsilon_z=\mu_z  =\Big(\frac{R_2}{R_2-R_1}\Big)^2\,\frac{r-R_1}{r}.
\end{equation}
  We refer to \cite{chen2009transformation,chen2010transformation} for a general framework for transformation optics.

It is free to set any values for the permeability and    permittivity  in the cloaked region: $r<R_1.$  Without loss of generality (cf.  \cite{Pendry.2006}),  we set
${\bs  \varepsilon}=\bs \mu={\bs I_3}$ for $r<R_1.$  

In what follows, we consider the transverse-electric (TE) polarised electromagnetic field, that is,  the electrical field only exists in the $z$ direction:  $\bs E=(0,0, u)^t.$ Then
by the first equation of   \eqref{mapped-Max},   we have
\begin{equation}\label{magenticH}
\bs H=(H_1, H_2, 0)^t=\frac 1 {\ri \omega \mu_0} \begin{cases}
\Big(\dfrac 1 {r-R_1} \dfrac{\partial u}{\partial\theta}, -\dfrac {r-R_1} r\dfrac{\partial u}{\partial r},0\Big)^t,\quad & {\rm if}\;\;   R_1<r<R_2,\\[12pt]
\Big(\dfrac 1  r\dfrac{\partial u}{\partial\theta},  -\dfrac{\partial u}{\partial r},0\Big)^t,\quad & {\rm if}\;\;   r<R_1\;\; {\rm or}\;\; r>R_2.
\end{cases}
\end{equation}
Eliminating $\bs H$ from the Maxwell  equations \eqref{mapped-Max},
we obtain the two-dimensional Helmholtz equations in polar coordinates:
\begin{align}
&{\mathcal L}_0[u]:=\frac{1}{r} \frac{\partial }{\partial r}\Big( r\frac{\partial u}{\partial r} \Big)+\frac{1}{r^2}\frac{\partial^2 u} {\partial \theta^2}+k^2 u=0, \qquad \; {\rm if}\;\;    r<R_1\;\; {\rm or}\;\; r>R_2;  \label{2DHelm2}\\[4pt]
&{\mathcal L}_1[u]:= \frac{1}{r-R_1} \frac{\partial }{\partial r}\Big((r-R_1) \frac{\partial u}{\partial r} \Big)+\frac{1}{(r-R_1)^2}\frac{\partial^2 u} {\partial \theta^2}+k^2 b^2u=0, \quad   {\rm if}\;\; R_1<r<R_2,\label{2DHelm1}
\end{align}
for all $\theta\in [0,2\pi),$ where
\begin{equation}\label{EkR}
 k=\omega\sqrt{\varepsilon_0\mu_0},\quad  b={R_2}/{(R_2-R_1)}.
\end{equation}
Conventionally,  we impose the Sommerfeld radiation boundary condition for the scattering wave: $u_{\rm sc}=u-u_{\rm in}$ (where
$u_{\rm in}$ is the  incident  wave, i.e., $\bs E_{\rm in}=(0,0, u_{\rm in})^t$, see Figure \ref{cloakA}):
\begin{equation}\label{somerfied}
\partial_r u_{\rm sc}-\ri k\, u_{\rm sc}=O(r^{-1/2}).
\end{equation}

\subsection{Exact  DtN boundary condition,   transmission conditions and CBCs}  We now consider  the boundary and transmission conditions to achieve perfect concealment of waves.

 Starting  with the outmost, we adopt the domain truncation by imposing an   artificial boundary condition at  $r=R_3>R_2$ using  the exact Dirichlet-to-Neumann (DtN) technique (see,  e.g., \cite{Gro.K95,Nede01}):
\begin{equation}\label{DtNbnd}
\partial_r u_{\rm sc}-{\mathcal T} _{R_3} u_{\rm sc} =0 \quad {\rm at}\;\;  r=R_3,
\end{equation}
where the DtN map  ${\mathcal T} _{R_3}$ is  defined as
\begin{equation}\label{DtNmap}
{\mathcal T} _{R_3} \psi = \Sum  \frac{k{\Hankel} '(kR_3)}{\Hankel (k_0 R_3)} \hat \psi_m  e^{\ri m\theta},\quad\hat \psi_m=\frac 1 {2\pi}\int_0^{2\pi} \psi(R_3,\theta) e^{-\ri  m\theta} d\theta,
\end{equation}
and $\Hankel$ is the Hankel function of the first kind.
This yields the exact artificial boundary conditions of  the total field:
\begin{equation}\label{uDtN}
\partial_r u-{\mathcal T} _{R_3} u = \partial_r u_{\rm in}-{\mathcal T} _{R_3} u_{\rm in}:=g   \quad {\rm at}\;\;  r=R_3.
\end{equation}
  {\bf Remark $2.1$.} {The exact DtN boundary condition is global in the physical space {\rm(}due to the involvement of  a Fourier series{\rm),} but it is local in the expansion coefficient space.  Given the  geometry of the cloak, we can fully take this advantage in both simulation and analysis.
\qed }
\vskip 6pt

For clarity of exposition, let us  denote
\begin{equation}\label{notations}
R_0=0;\;\;\; I_i= (R_{i}, R_{i+1}), \quad
  \Omega_{i}=I_i\times [0,2\pi),\quad  i=0,1,2;\quad \Omega=\bigcup_{i=0}^2 \Omega_i.
\end{equation}
Correspondingly, we define
\begin{equation}\label{HlayerA}
\bs H^{i}=\bs H|_{\Omega_i},\quad  \bs E^{i}=\bs E|_{\Omega_i}, \quad u^i=u|_{I_i},\quad   i=0,1,2.
\end{equation}

The conditions at the material  interface   $r=R_2$ are the standard transmission conditions, that is,
the  tangential components of  $\bs E$ and $\bs H$  are continuous across the interface (see,  e.g., \cite[Sec. 1.5]{orfanidis2002electromagnetic} and \cite{Monk03}):

\begin{equation}\label{Hcontinue}
\bs n\times (\bs E^{1} - \bs E^{2})=\bs 0, \quad \bs n\times (\bs H^{1} - \bs H^{2})=\bs 0\quad {\rm at}\;\; r=R_2,
\end{equation}
where $\bs n$ is the outer unit normal.   A direct calculation

from   \eqref{magenticH},  leads to
\begin{equation}\label{Hcongtiue}
 u^{1}-u^{2}=0, \quad
 b^{-1} \partial_r u^{1} -\partial_r u^{2}=0 \quad {\rm at}\;\; r=R_2.
\end{equation}

As mentioned in the introductory section, how to impose  suitable conditions so that there is no wave propagating into the cloaked region,  appears unsettled.  The analysis in Ruan et al. \cite{ruan2007ideal} implies that
\begin{equation}\label{cloakcondnot}
 \bs E^1\not =\bs 0,\quad \bs n\times (\bs E^{0} - \bs E^{1})\not=\bs 0 \quad {\rm at}\;\; r=R_1,
\end{equation}
while the tangential component is continuous across  the inner boundary, namely,
 \begin{equation}\label{cloakcondnotB}
 \bs n\times (\bs H^{0} - \bs H^{1})=\bs 0\quad {\rm at}\;\; r=R_1.
 \end{equation}
 Zhang et al. \cite{zhang2007response} demonstrates  that  the exotic physical effect  \eqref{cloakcondnot} is attributed to the surface  current
induced by the singular transformation.

We find from \eqref{magenticH} and \eqref{cloakcondnotB} that
\begin{equation}\label{R1cond}
\lim_{r\to R_1}\Big(\partial_r u^0-\frac{r-R_1}{r}\partial_r u^{1}\Big)=0,\quad
\end{equation}
which implies
\begin{equation}\label{R1cond2}
 \partial_r u^0(R_1)=0,\quad \partial_r u^1(R_1)\;\; \text{is finite}.
\end{equation}
{\bf Remark $2.2$.} {To shield the wave from propagating  into the cloaked region,   the PMC  condition {\rm(}i.e., $\bs n\times \bs H^1=\bs 0${\rm),}   is imposed  in many simulations in   Cartesian coordinates {\rm(}see,  e.g.,  \cite{cummer2006full,zhang2007response,Jichun.2012}{\rm).}  Unfortunately,  we infer  from \eqref{R1cond2}  that in the polar coordinates,   the {\rm PMC} condition {\rm(}equivalent to  $\partial_r u^1(R_1)$ being finite{\rm)}  does not lead to an independent condition. \qed}
\vskip 6pt


At this point, one condition is lacking at the inner boundary.   Our viewpoint is that the electromagnetic fields in the original coordinates  must still  be   finite after the coordinate transformation.
 Therefore, letting $r\to R_1^+$ in \eqref{magenticH}  yields
\begin{equation}\label{R1conds}
\frac{\partial u^{(1)}}{\partial \theta}(R_1,\theta)=0,\quad \theta\in [0,2\pi).
\end{equation}
We reiterate that  this condition can be viewed as the essential ``pole condition'', arisen from the  polar transformation (singular at the origin).     Note that  the ``pole'' condition is   imposed so that the solution in the polar coordinates has desired regularity, when it is transformed back to  Cartesian coordinates  (see, e.g., \cite{Got.O77, Boyd01}).
As shown in \cite{Shen97,ShenTaoWang2011},
the condition: $\partial_\theta u(0,\theta)=0$ is {\em essential}  for spectral-accurate computations.  Indeed, such a notion can be extended to  other singular transformations (see e.g., \cite{shen2009triangular}).
 In this context, the transformation \eqref{TransMap}  spans the origin to the circle $r=R_1,$  so the essential ``pole'' condition  is  transplanted to   $r=R_1^+.$

The problem  of interest is summarised  as follows:
\begin{align}
& {\mathcal L}_0[u^0]=0  \;\;\;  {\rm in}\;\;  \Omega_0;\quad\partial_\theta u^0(0,\theta)=\partial_r u^0(R_1,\theta)=0;   \label{domain0}\\
& {\mathcal L}_1[u^1]=0  \;\;\;  {\rm in}\;\;  \Omega_1;\quad\partial_\theta u^1(R_1,\theta)=0;   \label{domain1}\\
&  u^1=u^2,\quad b^{-1} \partial_r u^{1} =\partial_r u^{2}  \quad {\rm at}\;\; r=R_2, \label{transmit}\\
& {\mathcal L}_0[u^2]=0  \;\;\;  {\rm in}\;\;  \Omega_2;\quad \partial_r u^2-{\mathcal T} _{R_3} u^2 =g   \quad {\rm at}\;\;  r=R_3;  \label{domain2}
\end{align}
for all $\theta\in [0,2\pi).$  Note that  (i) the operators ${\mathcal L}_0$ and ${\mathcal L}_1$ are defined in \eqref{2DHelm2}-\eqref{2DHelm1};
(ii) the EPC: $\partial_\theta u^0(0,\theta)=0$ is imposed at the origin due to the singular polar transformation;
 and (iii) the source of the system is the incident wave in the data $g$ (cf. \eqref{uDtN}).

Remarkably, under the condition \eqref{R1cond}, the subproblem \eqref{domain0} is decoupled from \eqref{domain1}-\eqref{domain2}.   Moreover, we can show that  $u^0\equiv 0$ is the unique solution,  if the wave number $k$ is not
an eigenvalue of the  Bessel operator with the boundary conditions in \eqref{u0eqn}.  Indeed,  we write $u^0(r,\theta)=\sum_{|m|=0}^\infty \hat u_m^0(r)e^{\ri m\theta}.$
Then   \eqref{domain0}   reduces to
\begin{equation}\label{u0eqn}
\begin{split}
&\frac{1}{r} \frac{d }{dr}\Big( r\frac{d \hat u_m^0}{d r} \Big)-\frac{m^2
 }{r^2}\hat u_m^0+k^2 \hat u_m^0=0,\quad 0<r<R_1;\\
 & \hat u_m^0(0)=0,\;\; {\rm if}\;\; m\not=0;\quad \frac d {dr}\hat u_m^0(R_1)=0,\;\;\; m=0,\pm 1,\cdots.
 \end{split}
\end{equation}
We claim from the Sturm-Liouville theory of the Bessel operator (see,  e.g.,  \cite{Courant53,SLprb08}) the following conclusion.
\vskip 6pt
\noindent {\bf Proposition $2.1$.} {If $k$ is not an eigenvalue of the Bessel problem \eqref{u0eqn}, or equivalently,    $J_m'(kR_1)\not =0$ for any model $m, $   then we have $\hat u_m^0\equiv 0$ for every $m,$ so the solution of  \eqref{domain0} $ u^0\equiv 0.$ \qed}
\vskip 6pt


We see  that   with a reasonable assumption on the frequency of the incident wave,  the
cloak can perfectly shield the waves  from penetrating into the cloaked region.

\subsection{Fourier-Legendre-spectral-element method}\label{sect2}

We next present an accurate and efficient numerical  method for solving  \eqref{domain1}-\eqref{domain2}.  Observe from
 \eqref{uDtN} that the DtN boundary condition in \eqref{domain2} is global in the physical space, but it is local in the frequency   space of Fourier expansion. It is therefore advantageous to use  Fourier spectral approximation in $\theta$-direction, and Legendre spectral-element method in $r$-direction.

We  expand  the solution and given data in  Fourier series:
\begin{equation}\label{polarexpan1}
\big\{u^j(r,\theta),g(\theta)\}=\sum_{|m|=0}^\infty\big\{\hat u_m^j(r), \hat g_m\big\}e^{\ri m\theta},\quad j=1,2.
\end{equation}
Then  \eqref{domain1}-\eqref{domain2} reduce to a sequence of  one-dimensional equations:
\begin{align}
& \frac{1}{r-R_1} \frac{d}{dr}\Big((r-R_1) \frac{d \hat u_m^1}{d r} \Big)-\frac{m^2}{(r-R_1)^2}\hat u_m^1 +k^2 b^2\hat u_m^1=0, \quad  R_1<r<R_2; \label{1DHelm1A}\\
&\hat u_m^1(R_1)=0, \;\;{\rm if}\;\; m \neq 0;\quad \hat u_m^1(R_2)=\hat u_m^2(R_2), \quad
b^{-1} \frac d{dr} \hat u_m^1(R_2) =\frac  d{dr}\hat u_m^2(R_2);\label{1dtransmit}\\
&\frac{1}{r} \frac{d }{dr}\Big( r\frac{d \hat u_m^2}{d r} \Big)-\frac{m^2
 }{r^2}\hat u_m^2+k^2 \hat u_m^2=0,\quad R_2<r<R_3;\label{1DHelm1B}\\
& \Big(\frac d {dr} -{\mathcal T}_{m,k} \Big) \hat u_m^2(R_3)=\hat g_m, \quad \text{where}\;\; {\mathcal T}_{m,k}:=\frac{k{\Hankel} '(kR_3)}{\Hankel (k R_3)}. \label{1DHelm4D}
 \end{align}
 Note  that the global DtN boundary condition is decoupled for each mode $m,$ and by the property of the Hankel function, we have ${\mathcal T}_{-m,k}={\mathcal T}_{m,k}$ (see  \cite[(2.32)]{She.W07}).

 Hereafter, let $\Lambda=(a,b)$  and  $\varpi(x)>0$ be a generic weight function on $\Lambda,$ which is absolutely integrable.   Let $H^s_\varpi(\Lambda)$ be the weighted Sobolev space as defined in Admas \cite{Adams03}. In particular, $L^2_\varpi(\Lambda)=H^0_\varpi(\Lambda)$ with the inner product $(\cdot,\cdot)_\varpi$ and norm $\|\cdot\|_\varpi.$ We drop the weight function, whenever $\varpi=1.$

 Let $I_1=(R_1,R_2)$ and $I_2=(R_2,R_3)$ as before, and     let $I=(R_1, R_3).$
Define the weight function $\omega$ and the piecewise constant  function $\rho$:  
\begin{equation}\label{weightgfun}
\omega(r)=\begin{cases}
r-R_1,\quad & {\rm if}\;\; r\in I_1,\\
r,\quad &{\rm if}\;\; r\in I_2,
\end{cases}
\quad
 \rho =\begin{cases}
b^2,\quad & {\rm if}\;\; r\in I_1,\\
1,\quad &{\rm if}\;\; r\in I_2.
\end{cases}
\end{equation}
Recall that $b=R_2/(R_2-R_1)$ as defined in \eqref{EkR}.
 The weak form of \eqref{1DHelm1A}-\eqref{1DHelm4D} is to find $\hat u_m\in Y_m(I)$ for each mode $m,$ such that
\begin{equation}\label{Iweakform}
\begin{split}
{\mathcal B}_m(\hat u_m, v):=&(\hat u_m',  v')_\omega+m^2 (\hat u_m,  v)_{\omega^{-1}} -k^2
(\rho \hat u_m, v)_{\omega}\\
&-R_3 {\mathcal T} _{m,k}\, \hat u_m(R_3) \bar v(R_3)= R_3\hat g_m\bar v(R_3), \quad \forall v\in Y_m(I),
\end{split}
\end{equation}
where $\bar v$ is the complex conjugate of $v.$  We show in Appendix \ref{prop:uniquesolu} the following proposition on the  well-posedness
of   \eqref{Iweakform}.
\vskip 6pt
\noindent {\bf Proposition $2.2$.} For each mode $m,$ the problem \eqref{Iweakform} has a unique solution $\hat u_m\in Y_m(I).$
\vskip 6pt

We now discuss  the numerical solution of  \eqref{Iweakform}.
Let ${\mathbb P}_N$ be the complex-valued polynomials of degree at most $N,$ and
let $\bs N=(N_1,N_2).$  We introduce the  approximation space:
\begin{equation}\label{YmN}
Y_m^{\bs N}(I)=\big\{u\in Y_m(I)\,:\, u|_{I_i}\in{\mathbb P}_{N_i},\;\; i=1,2\big\},
\end{equation}
where the functions are continuous across $r=R_2$ (cf. \eqref{1dtransmit}).  Given a cut-off number $M>0,$ the FLSEM approximation to the solution of
\eqref{domain1}-\eqref{domain2} is
\begin{equation}\label{Fourierapp}
u_M^{\bs N}(r,\theta)=\sum_{|m|=0}^M \hat u_m^{\bs N}(r) e^{\ri m\theta}, \;\;\;
\end{equation}
where   $\{\hat u_m^{\bs N}\}$ are  computed from  the Legendre-spectral-element approximation to  \eqref{Iweakform}, that is,   find $\hat u_m^{\bs N}\in Y_m^{\bs N}(I) $  such that
\begin{equation}\label{IweakformNumer}
\begin{split}
{\mathcal B}_m(\hat u_m^{\bs N}, v)= R_3\hat g_m\bar v(R_3), \quad \forall v\in Y_m^{\bs N}(I),\quad 0\le |m|\le M.
\end{split}
\end{equation}
{\bf Remark $2.3$.} {We can apply the same argument as for Proposition $2.2$ to show that \eqref{IweakformNumer} admits a unique solution in $Y_m^{\bs N}(I)$. \qed }

For each mode $m,$ the two-domain spectral-element scheme
\eqref{IweakformNumer} can be implemented efficiently by using the modal Legendre polynomial basis and the Schur complement technique (cf. \cite{JiWuMaGuo}).  Here, we omit the details.
We provide in section \ref{sect:numer} ample numerical results to demonstrate that our proposed CBCs and FLSEM,   leads to  accurate simulations of the ideal circular cylindrical  cloak.

\section{Elliptic cylindrical cloaks}\label{sect:elliptic}

The elliptic cylindrical cloaks have been much less studied (see,   e.g.,  \cite{jiang2008arbitrarily,kwon2008two,cojocaru2009exact}), compared with intensive investigations of  the circular cylindrical cloaks. Though it is straightforward to extend the coordinate transformation in  \cite{Pendry.2006} to the elliptic case,  the coordinate transformation
possesses   quite different nature of singularity.  Thus, much care is needed to impose CBCs,  as to be shown shortly.

\subsection{Elliptic coordinates and Mathieu functions} To formulate the problem and algorithm, we briefly review the elliptic coordinates and the related angular and radial Mathieu functions.
The elliptic coordinates $(\xi,\eta)$ are related to the Cartesian coordinates ${\bs x}=(x, y)$ by
\begin{equation} \label{ellpco}
x=a \cosh \xi \cos \eta, \quad y=a \sinh \xi \sin \eta,\quad a>0,
\end{equation}
where $\xi\in [0,\infty)$ and  $\eta\in [0,2\pi).$  The coordinate lines  are confocal ellipses (of constant $\xi$)  and
 hyperbolae (of constant $\eta$) with foci fixed at $-a$ and $a$ on the $x$-axis.
The scale factors and the Jacobian of the elliptic coordinate system are
 \begin{equation}\label{scalesystem}
 h=h_\xi=h_\eta=a \sqrt {\cosh ^2 \xi -\cos ^2 \eta},\quad J=h_\xi h_\eta=h^2.
 \end{equation}

The (angular) Mathieu equation reads  (cf. \cite{Abr.S84}):

\begin{equation}\label{mathieu}
\frac{d^2\Phi}{d\eta^2}+(\lambda-2q\cos 2\eta)\Phi=0\;\; {\rm with}\;\;  q=\frac{a^2k^2} 4,
\end{equation}
where  $\lambda$ is the separation constant.
The angular Mathieu equation \eqref{mathieu}  supplemented with  periodic boundary conditions  admits
a countable set of eign-pairs : 
\begin{equation}\label{Phi}
\big\{\lambda_m^{\rm c}(q),\; \ce_m(\eta;q)\big\}_{m=0}^\infty, \quad \big \{\lambda_m^{\rm s}(q),\;  \se_m(\eta;q)\big\}_{m=1}^\infty.
\end{equation}
Note that the symbols  ``$\ce$'' and ``$\se$'',  abbreviation of ``cosine-elliptic"
and  ``sine-elliptic", were first introduced in \cite{Whi.W27}.
To facilitate the analysis afterwards, let us denote
\begin{equation}\label{newlambda}
\tilde{\lambda}_m^{\rm c}(q):=\lambda_m^{\rm c}(q)+2q, \quad \tilde{\lambda}_m^{\rm s}(q):=\lambda_m^{\rm s}(q)+2q.
\end{equation}
Then for any fixed  $q>0,$ from the standard Sturm-Liouville theory (cf.   \cite{Courant53}), the eigenvalues are in order of
\begin{equation}\label{fixedqeigen}
0<\tilde{\lambda}_0^{\rm c}(q)< \tilde{\lambda}_1^{\rm s}(q)<\tilde{\lambda}_1^{\rm c}(q)<\cdots < \tilde{\lambda}_m^{\rm s}(q)< \tilde{\lambda}_m^{\rm c}(q)<\cdots.
\end{equation}
When $q=0,$  the Mathieu
functions reduce
to the trigonometric functions:
\begin{equation}\label{q0q0}
\ce_0(\eta;0)=\frac 1 {\sqrt 2}; \quad \ce_m(\eta;0)=\cos(m\eta),\quad \se_m(\eta;0)=\sin(m\eta),\;\;\; m\ge 1,
\end{equation}
and correspondingly, $ \lambda_m^{\rm c}(0)=\lambda_m^{\rm s}(0)=m^2.$  Indeed, the angular Mathieu functions share many properties with their conterparts: cosines and sines.  For example,  $\ce_m(\eta; q)$ is an even function in $\eta,$ and $\se_m(\eta; q)$ is odd.
They are  $\pi$-periodic when $m$ is even, and $2\pi$-periodic when $m$ is odd.  Moreover,
the set of Mathieu functions $\{\ce_m, \se_{m+1}\}_{m=0}^\infty$ forms a complete orthogonal system
in  $L^2(0,2\pi)$ (cf.  \cite{mclachlan1964theory,Abr.S84}):
\begin{equation}\label{orth}
\int_0^{2\pi}\ce_m \,\ce_n\,d\eta=\int_0^{2\pi}\se_m\, \se_n\,d\eta
=\pi\delta_{mn};\quad \int_0^{2\pi}\ce_m\,\se_n\,d\eta=0.
\end{equation}

The radial  (or modified) Mathieu equation (cf. \cite{Abr.S84}):
\begin{equation}\label{mod_mathieu}
\frac{d^2 \Psi}{d \xi^2}-(\lambda-2q\cosh 2\xi)\Psi=0,
\end{equation}
plays an analogous role as
the Bessel equation in the  polar coordinates.
Like the Bessel functions, there are  several types of radial Mathieu functions, but  each type has
even and odd versions, quite different notation is used to denote such functions in literature \cite{mclachlan1964theory,Abr.S84}.
In this paper,  we adopt the notation and conventions in \cite{Abr.S84}, where
$\{{\rm Mc}_m^{(i)}; {\rm Ms}_m^{(i)}\}, i=1,2,3,4,$ correspond to the Bessel functions:
$J_m, Y_m, H_m^{(1)}, H_m^{(2)}$ in \cite{watson}, respectively.  In what follows, we shall just  use
the radial Mathieu functions of the first kind $\{{\rm Mc}_m^{(1)}(\xi;q); {\rm Ms}_m^{(1)}(\xi;q)\}$, and the
Mathieu-Hankel functions: $\{{\rm Mc}_m^{(3)}(\xi;q); {\rm Ms}_m^{(3)}(\xi;q)\}.$ Both types satisfy
\eqref{mod_mathieu}
with $\lambda=\lambda_m^{\rm c}$ and $\lambda=\lambda_m^{\rm s}$  for ${\rm Mc}_m$ and ${\rm Ms}_m,$ respectively.

\subsection{Maxwell equations  for ideal elliptic cylindrical cloak} Following the idea of  Pendry et al.
\cite{Pendry.2006}, a  coordinate transformation,  which  compresses the elliptic region $0\le \zeta<\xi_2$
into the elliptic annular region
 $0<\xi_1<\xi<\xi_2,$  was extended  to devise an elliptic cylindrical cloak
  (see e.g., \cite{ma2008material,cojocaru2009exact}):
\begin{equation}\label{TransMapEllip}
\xi=
 \frac {\zeta} d +\xi_1,\quad \eta=\eta, \quad z=z \;\;\; {\rm with}\;\;\; d=\frac {\xi_2} {\xi_2-\xi_1},
\end{equation}
where $(\zeta,\eta, z)$ is the elliptic-cylindrical  coordinates in the original space, and
$(\xi,\eta, z)$ is the coordinates  of the transformed space.
This leads to the study of the time-harmonic Maxwell  equations in the transformed space
with new material parameters: 
\begin{equation} \label{mapped-Max_Ellip}
\nabla \times {\bs E}-\ri \omega \mu_0 {\bs \mu} {\bs H}=0,\quad \nabla \times {\bs H} + \ri \omega \varepsilon_0 {\bs  \varepsilon} {\bs E} =0, \;\;\; \xi>\xi_1,
\end{equation}
where we have
\vspace*{6pt}
\begin{equation}\label{EllipMedia}
\begin{split}
&{\bs  \varepsilon}={\bs  \mu}={\rm diag}(\varepsilon_\xi,\varepsilon_\eta,\varepsilon_z),\;\; {\rm if}\;\; \xi_1<\xi<\xi_2;\quad {\bs  \varepsilon}={\bs  \mu}=\bs I_3, \;\; {\rm if}\;\; \xi>\xi_2,
\end{split}
\end{equation}
with the components in the cloak (cf. \cite{ma2008material}),  given by
\begin{equation}\label{cloakpara}
\varepsilon_{\xi}=\mu_\xi= \frac{1}{d},\quad  \varepsilon_\eta=\mu_\eta=d,  \quad \varepsilon_z=\mu_z  =d \frac{\cosh ^2(d(\xi-\xi_1))-\cos^2 \eta}{\cosh^2 \xi-\cos ^2 \eta}.
\end{equation}

As with the circular case, we consider the transverse-electric (TE) polarised electromagnetic field with  $\bs E=(0,0, v)^t.$  Like \eqref{magenticH},  the magnetic field in this context takes the form:
\begin{equation}\label{magenticH_Ellip}
\bs H=(H_1, H_2, 0)^t=\frac 1 {\ri \omega \mu_0} \begin{cases}
\Big(\dfrac {d} {h} \dfrac{\partial v}{\partial\eta}, -\dfrac {1}{d h}\dfrac{\partial v}{\partial \xi},0\Big)^t,\quad & {\rm if}\;\;   \xi_1<\xi<\xi_2,\\[12pt]
\Big(\dfrac 1  h\dfrac{\partial v}{\partial\eta},  -\dfrac 1 h \dfrac{\partial v}{\partial \xi},0\Big)^t,\quad & {\rm if}\;\;   \xi>\xi_2\;\; {\rm or}\;\; 0< \xi<\xi_1.
\end{cases}
\end{equation}
With the above polarisation, we eliminate $\bs H$ and obtain the Helmholtz equations, together with  exact DtN boundary at the outer ellipse $\xi=\xi_3(>\xi_2),$ in elliptic coordinates:
\begin{subequations}\label{2DHelm_Ellip}
\begin{gather}
 \big({d^{-2}} \partial^2_{\xi} +\partial ^2 _{\eta}\big)v +k^2 a^2 \big(\cosh^2(d(\xi-\xi_1))-\cos^2 \eta\big)v=0 \quad\;   {\rm in}\;\; \Lambda_1;\label{2DHelm_Ellip1}\\[4pt]
\big(\partial^2_{\xi} +\partial ^2 _{\eta}\big)v +k^2 a^2 \big(\cosh^2 \xi-\cos^2 \eta \big)v=0 \qquad\qquad  {\rm in}\;\;    \Lambda_0\cup\Lambda_2;  \label{2DHelm_Ellip2}
\\[4pt]
(\partial_{\xi} -{\mathbb T} _{\xi_3}) v=\phi   \quad\;\;  {\rm at}\;\;  \xi=\xi_3, \label{ellipticexact}
 \end{gather}
\end{subequations}
where
$$k=\omega\sqrt{\varepsilon_0\mu_0};\quad \Lambda_i:=(\xi_i,\xi_{i+1})\times [0,2\pi),\;\;\;  i=0,1,2\;\; {\rm with}\;\; \xi_0:=0,$$
and
${\mathbb T} _{\xi_3}$ is the DtN map (cf. \cite{Gro.K95,FSW09}), given by
\begin{equation}\label{DtNmap_Ellip}
\begin{split}
&{\mathbb  T} _{\xi_3} v= \sum_{m=0}^{\infty}  \frac{\partial_\xi {\rm Mc}_m^{(3)}(\xi_3;q)} {{\rm Mc}_m^{(3)}(\xi_3;q)} \hat v^{\rm c}_m(\xi_3)\, {\rm ce}_m(\eta;q)+\sum_{m=1}^{\infty}  \frac{\partial_\xi {\rm Ms}_m^{(3)}(\xi_3;q)} {{\rm Ms}_m^{(3)}(\xi_3;q)} \hat v^{\rm s}_m(\xi_3)\, {\rm se}_m(\eta;q),
\end{split}
\end{equation}
with
\begin{equation}\label{psic}
\hat v_m^{\rm c}(\xi_3)=\frac 1 {\pi}\int_0^{2\pi} v(\xi_3,\eta) \,{\rm ce}_m(\eta;q)\, d\eta, \quad \hat v_m^{\rm s}(\xi_3)=\frac 1 {\pi}\int_0^{2\pi} v(\xi_3,\eta) \,{\rm se}_m(\eta;q)\, d\eta.
\end{equation}
Note that in \eqref{ellipticexact},  $\phi$ is induced  by the incident wave, i.e.,  $ \phi=(\partial_{\xi} -{\mathbb T} _{\xi_3}) v_{\rm in}.$

Naturally,  we   impose continuity of  the tangential components of  $\bs E $ and $\bs H$ across  the elliptic interface $\xi=\xi_2,$ leading to  the transmission conditions as with \eqref{Hcongtiue}:
\begin{equation}\label{Hcongtiue_Ellip}
  v^{1}=v^{2}, \quad   d^{-1} \partial_{\xi} v^{1} =\partial_{\xi} v^{2},\quad {\rm at}\;\; \xi=\xi_2,
\end{equation}
where for clarity, we denote  $v^i=v|_{\Lambda_i}$ for $i=0,1,2.$

The critical issue is the imposition of CBCs at the inner boundary $\xi=\xi_1.$
%
To tackle this,  we  decompose the solution and data into {\rm ce}-  and  {\rm se}-components  as follows:
 \begin{equation}\label{polarexpan1_Ellip}
 \begin{split}
\big\{v;\, \phi\}&=\sum_{m=0}^\infty\big\{ \hat v^{\rm c}_m(\xi);\,  \hat\phi^{\rm c}_m\big\}{\rm ce}_m(\eta;q)+
\sum_{m=1}^\infty\big\{ \hat v^{\rm s}_m(\xi);\, \hat\phi^{\rm s}_m\big\}{\rm se}_m(\eta;q)
\\&:=
\big\{v_{\rm ce};\,\phi_{\rm ce}\}+\big\{v_{\rm se};\,\phi_{\rm se}\}.
\end{split}
\end{equation}
 Correspondingly, the polarized $\bs E$ and $\bs H$ fields
 are  split into  two components: $\bs E=\bs E_{\ce}+\bs E_{\se}$ and
$\bs H=\bs H_{\rm ce}+\bs H_{\rm se}.$  Following  the analytic study in  \cite{cojocaru2009exact}
and the $\delta$-perturbation analysis in  \cite{ruan2007ideal},  it is necessary to require   the   tangential component of $\bs H_{\ce}$ and   $\bs E_{\se}$ continuous  across the inner boundary $\xi=\xi_1$, leading to
\begin{equation}\label{Hcongtiue_Ellip2A}
  v^{0}_\se=v^{1}_\se, \quad   d^{-1} \partial_{\xi} v^{0}_\ce =\partial_{\xi} v^{1}_\ce\quad {\rm at}\;\; \xi=\xi_1.
\end{equation}
Nevertheless, we are short of one condition for each component.
Like the derivation of \eqref{R1conds} in the circular case,  we require  the fields in the  original space to be still   finite in the transformed space, leading to
  the essential ``pole'' conditions. 
  since the singularity of the transformation \eqref{TransMapEllip} only occurs at two points $(\xi_1,0)$
and $(\xi_1,\pi),$ as opposite to the  circular case, taking   limit  $\xi\to \xi_1^+$ in \eqref{magenticH_Ellip}, leads to
\begin{equation}\label{R1conds_Ellip}
\partial_{\xi}v^1(\xi_1,0) = \partial_{\xi}v^1(\xi_1,\pi)=0,\quad
\partial_{\eta}v^1(\xi_1,0)=\partial_{\eta}v^1(\xi_1,\pi) =0.
\end{equation}
Using the property  (cf.  \cite{Abr.S84}):
\begin{equation}\label{propty}
 \ce_m'(0;q)= \ce_m'(\pi;q)=0,\quad \se_m(0;q)=\se_m(\pi;q)=0,
\end{equation}
 we find from \eqref{polarexpan1_Ellip} that \eqref{R1conds_Ellip} is equivalent to
\begin{equation}\label{R1conds_EllipA2}
\partial_{\xi}v^1_\ce(\xi_1,0) =0,\quad
v^1_\se(\xi_1,0) =0.
\end{equation}

We summarize the  problem that models   the ideal elliptic cylindrical cloak:  given $\phi=\phi_\ce+\phi_\se, $ find
\begin{equation}\label{vformv}
v=v_\ce+v_\se\;\; {\rm with}\;\;   v^i=v_\ce^i+v_\se^i= (v_\ce+v_\se)|_{\Lambda_i},\;\;\; i=0,1,2,
\end{equation}
satisfying the following systems.
\begin{itemize}
\item[(i)] For the $\ce$-component in the cloaked region $\Lambda_0$:
\begin{align}
 &\big(\partial^2_{\xi} +\partial ^2 _{\eta}\big)v_\ce^0 +k^2 a^2 \big(\cosh^2\xi-\cos^2 \eta\big)v_\ce^0=0, \quad 0<\xi<\xi_1; \label{zerocomponent}\\
 & \partial_{\xi}v^0_\ce(0,0)= \partial_{\xi}v^0_\ce(\xi_1,0)=0. \label{pole-ellipse}
\end{align}
Note that the essential pole condition at the origin is  necessary,  while  $\partial_{\xi}v^0_\ce(\xi_1,0)=0 $
is derived from the second condition in  \eqref{Hcongtiue_Ellip2A} and \eqref{R1conds_EllipA2}.
\item[(ii)] For the $\ce$-component in  $\Lambda_1\cup\Lambda_2$:
\begin{subequations}\label{2DHelm_Ellip2A}
\begin{gather}
 \big(d^{-2} \partial^2_{\xi} +\partial ^2 _{\eta}\big)v_\ce^1 +k^2 a^2 \big(\cosh^2(d(\xi-\xi_1))-\cos^2 \eta\big)v_\ce^1=0 \quad\;   {\rm in}\;\; \Lambda_1;\label{2DHelm_Ellip12A}\\[4pt]
\partial_{\xi} v^1_\ce(\xi_1,0) =0;\quad\;\;    v^{1}_\ce=v^{2}_\ce, \quad   d^{-1} \partial_{\xi} v^{1}_\ce =\partial_{\xi} v^{2}_\ce \qquad {\rm at}\;\;\; \xi=\xi_2;\label{transmitAA}
\\[4pt]
\big(\partial^2_{\xi} +\partial ^2 _{\eta}\big)v_\ce^2 +k^2 a^2 \big(\cosh^2 \xi-\cos^2 \eta \big)v_\ce^2=0 \qquad\qquad \;\;\;\; {\rm in}\;\;    \Lambda_2,   \label{2DHelm_Ellip22A}
\\[4pt]
(\partial_{\xi} -{\mathbb T} _{\xi_3}) v_\ce^2=\phi_\ce   \qquad\;\;  {\rm at}\;\;  \xi=\xi_3. \label{ellipticexact2A}
 \end{gather}
\end{subequations}
\item[(iii)]  The $\se$-component $v_\se$ satisfies the same equations in (i)-(ii)  with $v_\se^i$ and  $\phi_\se$
in place of $v_\ce^i$ and  $\phi_\ce,$ respectively, while   \eqref{pole-ellipse} and the first condition in
\eqref{transmitAA} are  respectively  replaced by
\begin{equation}\label{senewcond}
 v^0_\se(0,0)= v^0_\se(\xi_1,0)=0;\quad  v^1_\se(\xi_1,0) =0.
 \end{equation}
\end{itemize}

We see that  $v_\ce^0$ and $v_\se^0$ are decoupled from $v_\ce^i$ and $v_\se^i, \; i=1,2$.  Indeed,
using \eqref{mathieu} and  the expansion  \eqref{polarexpan1_Ellip},   the problem \eqref{zerocomponent}-\eqref{pole-ellipse} reduces to
\begin{align}
 &(\hat v_m^{\rm c})''(\xi) -\big(\lambda_m^{\rm c}-2q \cosh (2\xi)\big)\, \hat v_m^{\rm c}(\xi)=0, \quad 0<\xi<\xi_1, \label{zerocomponent2A}\\
 & (\hat v_m^{\rm c})'(0)=(\hat v_m^{\rm c})'(\xi_1)=0, \label{pole-ellipse2A}
\end{align}
and similarly, we have
\begin{align}
 &(\hat v_m^{\rm s})''(\xi) -\big(\lambda_m^{\rm s}-2q \cosh (2\xi)\big)\, \hat v_m^{\rm s}(\xi)=0, \quad 0<\xi<\xi_1, \label{zerocomponent2As}\\
 & \hat v_m^{\rm s}(0)=\hat v_m^{\rm s}(\xi_1)=0. \label{pole-ellipse2As}
\end{align}
Similar to  Proposition $2.1$,  one deduces  from the standard theory of ordinary differential equations
(cf. \cite{Courant53,SLprb08}) and also from the  properties of
Mathieu functions (cf. \cite{Abr.S84}) the following conclusions.
\vskip 6pt
\noindent {\bf Proposition $3.1.$} {If  $k$ and $\xi_1$ are chosen such that
  ${{\rm Mc}_m^{(1)}}'(k\xi_1)\not =0$ and ${{\rm Ms}_m^{(1)}}(k\xi_1)\not =0$
   for any mode $m,$ then we have $\hat v_m^{\rm c}=\hat v_m^{\rm s}\equiv 0$ for every $m,$ so the problem \eqref{zerocomponent}-\eqref{pole-ellipse} and the problem of   ${\se}$-component in $\Lambda_0$ with
   \eqref{senewcond},
     both have only trivial solutions in the cloaked region. \qed}
\vskip 6pt

\subsection{Mathieu-Legendre-spectral-element method}
In what follows,  we present an accurate and efficient numerical algorithm to simulate the ideal elliptic cylindrical  cloak.

Using the Mathieu expansion in $\eta$-direction (see  \eqref{polarexpan1_Ellip}), we obtain the system of   the $\ce$-component in $\Lambda_1\cup\Lambda_2$:
\begin{subequations}\label{1Dce}
\begin{align}
&-{d^{-2}}(\hat v_m^{\rm c})''(\xi)+\big(  \tilde {\lambda}_m^{\rm c}-4q \cosh^2(d(\xi-\xi_1)) \big) \hat v_m^{\rm c}=0,\quad \xi_1<\xi<\xi_2; \label{1DceA} \\[4pt]
&(\hat v_m^{\rm c})'(\xi_1)=0; \quad \hat v_m^{\rm c}(\xi_2^-)=\hat v_m^{\rm c}(\xi_2^+),\;\; d^{-1}(\hat v_m^{\rm c})'(\xi_2^-)=(\hat v_m^{\rm c})'(\xi_2^+);\label{1DceB}\\[4pt]
&-(\hat v_m^{\rm c})''(\xi)+\big(\tilde {\lambda}_m^{\rm c}-4q\cosh^2(\xi)\big)\hat v_m^{\rm c}(\xi)=0,\quad \xi_2<\xi<\xi_3; \label{1DceC}\\[4pt]
&\Big(\frac{d}{d\xi}-{\mathcal D}_m^{\rm c}\Big)\hat v_m^{\rm c}(\xi_3)=\hat \phi_m^{\rm c},\quad \text{where}\;\;{\mathcal D}_m^{\rm c}:=\frac{{{\rm Mc}_m^{(3)}}'(\xi_3;q)}{{\rm Mc}_m^{(3)}(\xi_3;q)}.\label{1DceD}
\end{align}
\end{subequations}
Recall that  $\tilde {\lambda}_m^{\rm c}:=\lambda_m^{\rm c}+2q >0$ (see \eqref{newlambda}).
The $\se$-component  satisfies the same system with $\hat v_m^{\rm s}$,  $\hat \phi_m^{\rm s}$ and $\tilde  {\lambda}_m^{\rm s}$
in place of $\hat v_m^{\rm c}$, $\hat \phi_m^{\rm c}$ and $\tilde  {\lambda}_m^{\rm c}$ in \eqref{1Dce},  respectively, while  the first condition in
\eqref{1DceB}  and  ${\mathcal D}_m^{\rm c}$  in \eqref{1DceD}, are respectively  replaced by
\begin{equation}\label{senewcond2A}
\hat v_m^{\rm s}(\xi_1)=0,   \quad  {\mathcal D}_m^{\rm s}:=\frac{{{\rm Ms}_m^{(3)}}'(\xi_3;q)}{{\rm Ms}_m^{(3)}(\xi_3;q)},\quad m\ge 1.
 \end{equation}

With a little abuse of notation, we still denote $I_1=(\xi_1,\xi_2), $ $I_2=(\xi_2,\xi_3),$ and  $I=(\xi_1,\xi_3)$.
To formulate the problem into a compact form (see \eqref{EllipWeakc} below), we introduce the piecewise  functions:  
\begin{equation} \label{EllipWeight}
\varpi=\begin{cases}
d^{-1},\quad & {\rm if}\;\; \xi \in I_1,\\
1,\quad &{\rm if}\;\; \xi \in I_2,
\end{cases}
\quad
 \chi=\begin{cases}
\cosh^2(d(\xi-\xi_1)),\quad & {\rm if}\;\; \xi \in I_1,\\
\cosh^2 \xi,\quad &{\rm if}\;\; \xi\in I_2.
\end{cases}
\end{equation}
Recall that $d=\xi_2/(\xi_2-\xi_1)$  defined in \eqref{TransMapEllip}.
Apparently, these two functions are uniformly bounded.

 The weak form of \eqref{1Dce} is to find $\hat v_m^{\rm c}\in H^1(I)$ for each mode $m,$ such that
\begin{equation}\label{EllipWeakc}
\begin{split}
{\mathcal B}_m^{\rm c}(\hat v_m^{\rm c}, \psi):=&(\varpi (\hat v_m^{\rm c})',  \psi')+\tilde {\lambda}_m^{\rm c} (\varpi^{-1}\hat v_m^{\rm c},  \psi) -4q
(\varpi^{-1}\chi\hat v_m^{\rm c}, \psi)\\
&-{\mathcal D} _{m}^{\rm c} \hat v_m^{\rm c}(\xi_3) \bar \psi(\xi_3)= \hat \phi_m^{\rm c}\bar \psi(\xi_3), \quad \forall \psi \in H^1(I), \;\;m=0, 1, \cdots.
\end{split}
\end{equation}
Similarly, the weak form of  the ${\rm se}$-component  is to find $\hat v_m^{\rm s}\in {}_0H^1(I):=\{  v\in H^1(I)\;:\; v(\xi_1)=0       \}$ for each mode $m,$ such that
\begin{equation}\label{EllipWeaks}
{\mathcal B}_m^{\rm s}(\hat v_m^{\rm s}, \psi)=\hat \phi_m^{\rm s}\bar \psi(\xi_3), \quad \forall \psi \in {}_0H^1(I),\;\; m=1, 2,\cdots,
\end{equation}
where the bilinear form ${\mathcal B}_m^{\rm s}(\cdot,\cdot)$ is defined by replacing  $\tilde {\lambda}_m^{\rm c}$ and
${\mathcal D}_m^{\rm c}$ in ${\mathcal B}_m^{\rm c}(\cdot,\cdot)$ by   $\tilde {\lambda}_m^{\rm s}$ and
${\mathcal D}_m^{\rm s},$  respectively.

Like Proposition $2.2$,   we next show the unique solvability of  \eqref{EllipWeakc} and \eqref{EllipWeaks}.  We postpone its proof in Appendix \ref{AppendixA}.

\vskip 6pt
\noindent {\bf Proposition $3.2$.}  {For each mode $m,$ the problem \eqref{EllipWeakc} {\rm (}resp. \eqref{EllipWeaks}{\rm)}
 has a unique solution $\hat v_m^{\rm c}\in H^1(I)$ {\rm(}resp. $\hat v_m^{\rm s}\in {}_0H^1(I)${\rm).} \qed}
\vskip 6pt

We now introduce the numerical schemes.  Define  the approximation spaces:
\begin{equation}\label{EllipApprox}
Z_{m}^{\rm c, \bs N}(I)=\big\{v \in H^1(I)\,: v|_{I_i}\in {\mathbb P}_{N_i},\;\;i=1,2\big\}; \quad Z_{m}^{\rm s,\bs N}(I)=Z_{m}^{\rm c,\bs N}(I)\cap H_0^1(I),
\end{equation}
where $\bs N=(N_1, N_2).$
 Given a cut-off number $M>0$, the MLSEM  approximation to the solution of \eqref{2DHelm_Ellip} in $\Lambda_1 \cup \Lambda_2$ is
\begin{equation}\label{Mathieuapp}
v_M^{\bs N}(\xi,\eta)=\sum_{m=0}^{M} \hat v_{m}^{\rm c,\bs N}(\xi) \ce_m(\eta; q)+\sum_{m=1}^{M}
 \hat v_{m}^{\rm s,\bs N}(\xi) \se_m(\eta; q),
\end{equation}
and   $\{ \hat v_{m}^{\rm c,\bs N},   \hat v_{m}^{\rm s,\bs N}\}$ are computed from    the  Legendre-spectral-element schemes:~
find $\hat v_{m}^{\rm c,\bs N}\in Z_{m}^{\rm c,\bs N}(I) $ such that
\begin{equation}\label{EllipMathieuapp}
{\mathcal B}_{m}^{\rm c}(\hat v_{m}^{\rm c ,{\bs N}},\psi)=\hat \phi_m^{\rm c} \bar \psi(\xi_3), \quad \forall \psi \in
Z_{m}^{\rm c, \bs N}(I), \quad 0\le m\le M,
\end{equation}
and
find $\hat v_{m}^{\rm s,\bs N}\in Z_{m}^{\rm s,\bs N}(I) $ such that
\begin{equation}\label{EllipMathieuapp2A}
{\mathcal B}_{m}^{\rm s}(\hat v_{m}^{\rm s ,{\bs N}},\psi)=\hat \phi_m^{\rm s} \bar \psi(\xi_3), \quad \forall \psi \in
Z_{m}^{\rm s, \bs N}(I), \quad 1\le m\le M.
\end{equation}
\vskip 6pt
 {\bf Remark $3.1$.} {The unique solvability of  \eqref{EllipMathieuapp}-\eqref{EllipMathieuapp2A} can be shown as the continuous problems in
Proposition $3.2$.  \qed}
\vskip 6pt

As with the circular case, the two-domain spectral-element scheme for each mode can be implemented  by using the modal Legendre polynomial basis and the Schur complement technique (cf. \cite{JiWuMaGuo}), but it is noteworthy that the resulted linear systems are full and dense  due to the involvement of the non-polynomial weight function $\chi$ in \eqref{EllipWeight}.


\section{Numerical results}\label{sect:numer}

In this section,  we provide ample numerical results to  demonstrate that our proposed approach produces accurate simulation of the ideal circular and elliptic  cylindrical cloaks.

\subsection{Circular cylindrical  cloaks}
Assuming that the incident  wave is a  plane  wave with an incident angle $\theta_0$:
 \begin{equation}\label{uinplane}
u_{{\rm in}}(r,\theta)=e^{{\rm i}k r \cos(\theta-\theta_0)}=\Sum {\rm i}^m {J}_m(kr) e^{{\rm i }m (\theta-\theta_0)},
\end{equation}
 we can derive from  the full-wave analysis in Ruan et al.  \cite{ruan2007ideal}  that  the ideal cloaking problem   admits  the exact solution:
\begin{equation}\label{uexactA}
u(r,\theta)
=\begin{cases}
u_{{\rm in}}(b(r-R_1),\theta) ,\quad  & {\rm if}\;\; R_1< r <R_2,\\
u_{{\rm in}}(r,\theta) ,\quad  & {\rm if}\;\;  r> R_2,
\end{cases}
\end{equation}
which vanishes in the cloaked region: $r<R_1.$


We first  examine the  numerical error: 
$E_{\bs N}= \smash{\max_{|m|\leq M}} \|\hat u_m-\hat u_m^{\bs N}\|_{\bs N, \infty}, $
where $\|\cdot\|_{\bs N, \infty}$ denotes the maximum pointwise errors at the Legendre-Gauss-Lobatto points (with a linear transformation) used in each subinterval. In the computation,   we take $M=70$ (so that  the truncation  error in $\theta$ direction  is negligible), and  choose  $\theta_0=0$ and $(R_1, R_2, R_3)=(0.2, 0.6, 1.0).$ In Figure \ref{Highfreq} (left), we plot $\log_{10}(E_{\bs N})$ against $\bs N=(N,N)$ for  $k=30,50,70.$   Observe that  the error decays exponentially, when
   $N>N_0(k).$  The expected transition value  $N_0(k)$ can be estimated by using  the notion of ``number-of-points-per-wavelength'' (cf. \cite{Got.O77}). Indeed, for large $k,$
   the Bessel function behaves like (cf. \cite{Abr.S84}):
   $$J_m(z)\sim  \sqrt{\frac {2}{\pi z}}\cos\Big(z-\frac{m \pi} 2-\frac \pi 4  \Big). $$
   We infer from the approxibility of Legendre polynomial expansions  to trigonometric functions (cf. \cite{Got.O77}) that  as soon as
   $$N>\frac {e k} 4 \max\{R_2, R_3-R_2 \}-\frac 1 2,$$
   the error begins to decay.  Approximately,   we take $N_0(k)$ to be ceiling round-off of this low bound.
For $k=30,50, 70,$  we find that $N_0=12,20, 29,$ respectively, which agrees with the numerical results in Figure \ref{Highfreq} (left).
In Figure \ref{Highfreq} (right), we plot the zeroth mode in the expansion \eqref{uinplane}  (see the solid line) versus the   numerical approximation of  $\hat u_0^{\bs N}(r)$  with
$k=70$ and ${\bs N}=(50,50) $ (with marker  ``{\tiny ${}^+$}'').  We see that this mode  is not continuous across  the inner boundary $r=R_1,$   which  is the major reason for the  surface currents (cf. \cite{zhang2007response}) and the violation of PEC condition (cf. \eqref{cloakcondnot}).
\begin{figure}[h!]
  \centering
    \includegraphics[width=0.49\textwidth]{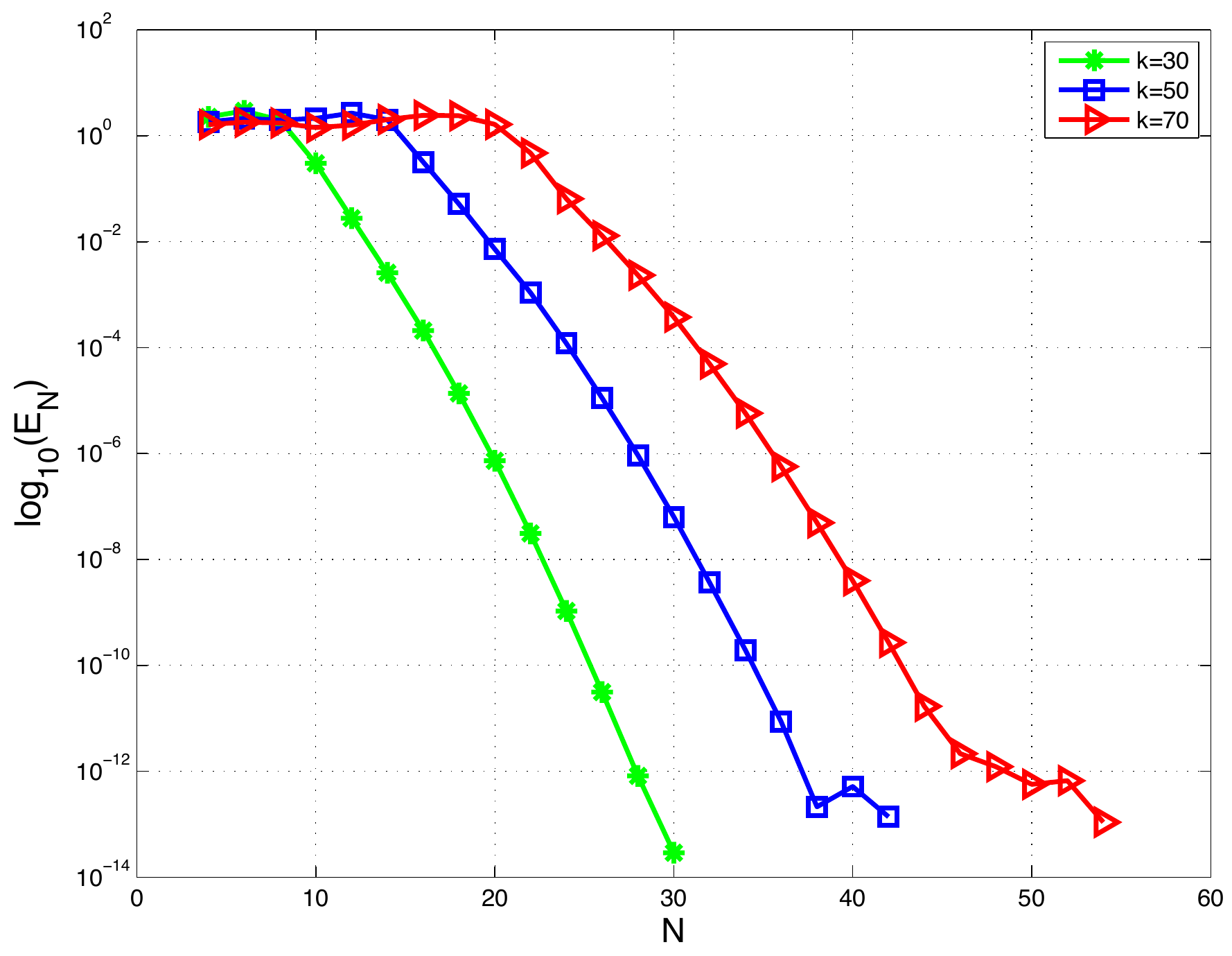}\;\;
    \includegraphics[width=0.45\textwidth]{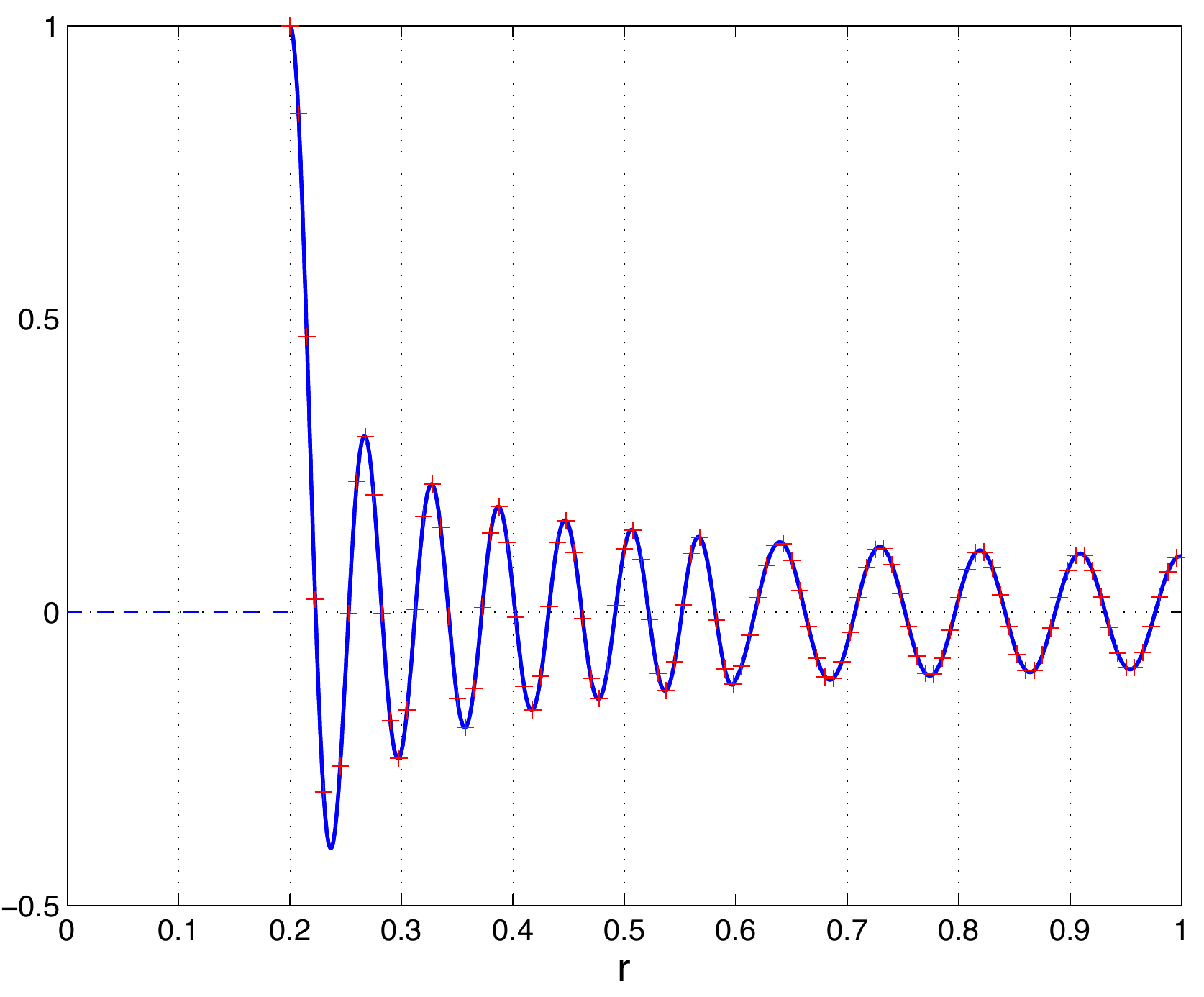}
   \caption{Left: errors  against various $N$ for some samples of  $k$. Right:  the zeroth mode (in solid line) versus its  numerical approximation $\hat u_0^{\bs N}$ (marked  by  ``{\tiny ${}^+$}'',  with  $k=70$ and ${\bs N}=(50,50)$). }
    \label{Highfreq}
\end{figure}

We next illustrate the electric wave propagations and profiles under different incident angles and frequencies.
In  Figure \ref{cloakk20},  we depict the electric-field distributions (real and imaginary parts in the top row) simulated by the proposed FLSEM with  $\theta_0=0, k=20, $    $(R_1, R_2, R_3)=(0.2, 0.6, 1.0),$
$M=25$ and $\bs N=(20,20)$.
 We see that when a TE plane  wave is incident  on the circular cloak,  it is completely guided and bent around the cloaked region without inducing any scattering waves. Moreover,  the propagating wavefronts perfectly emerge from  the other side of the cloaked region without any distortion,
 which are best  testified to  by profiles of   ${\rm Re}\{u_{M}^{\bs N}\}$ and  ${\rm Im}\{u_{M}^{\bs N}\}$  along $x$-axis (cf. \eqref{Fourierapp}) in  Figure \ref{cloakk20}.    Once again, we observe that
 the real part is discontinuous across the inner boundary,  attributed to the surface currents induced by the singular coordinate transformation (cf. \cite{zhang2007response}).
\begin{figure}[h!]
  \centering
    \includegraphics[width=0.49\textwidth]{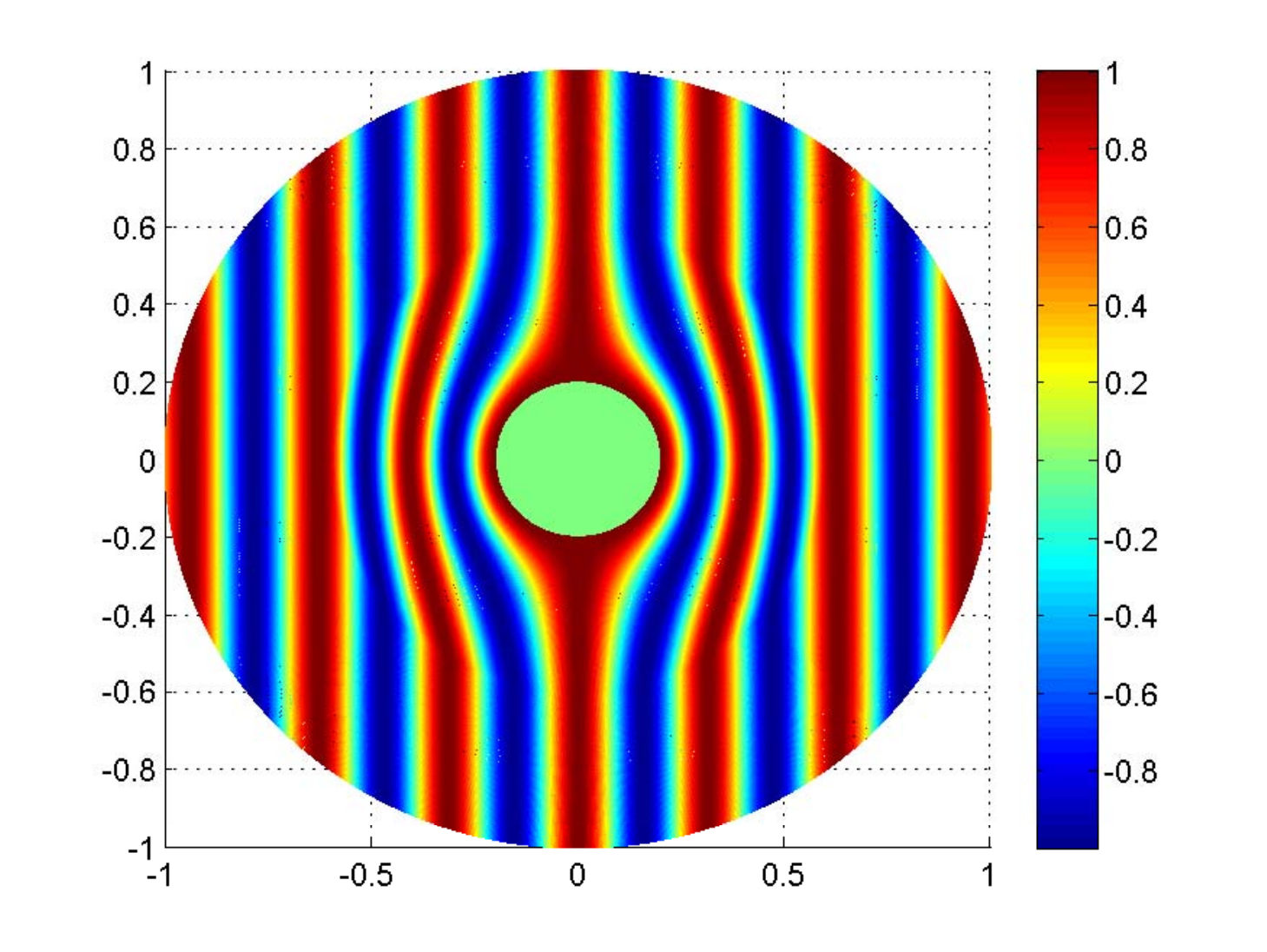}\hspace*{-18pt}
    \includegraphics[width=0.49\textwidth]{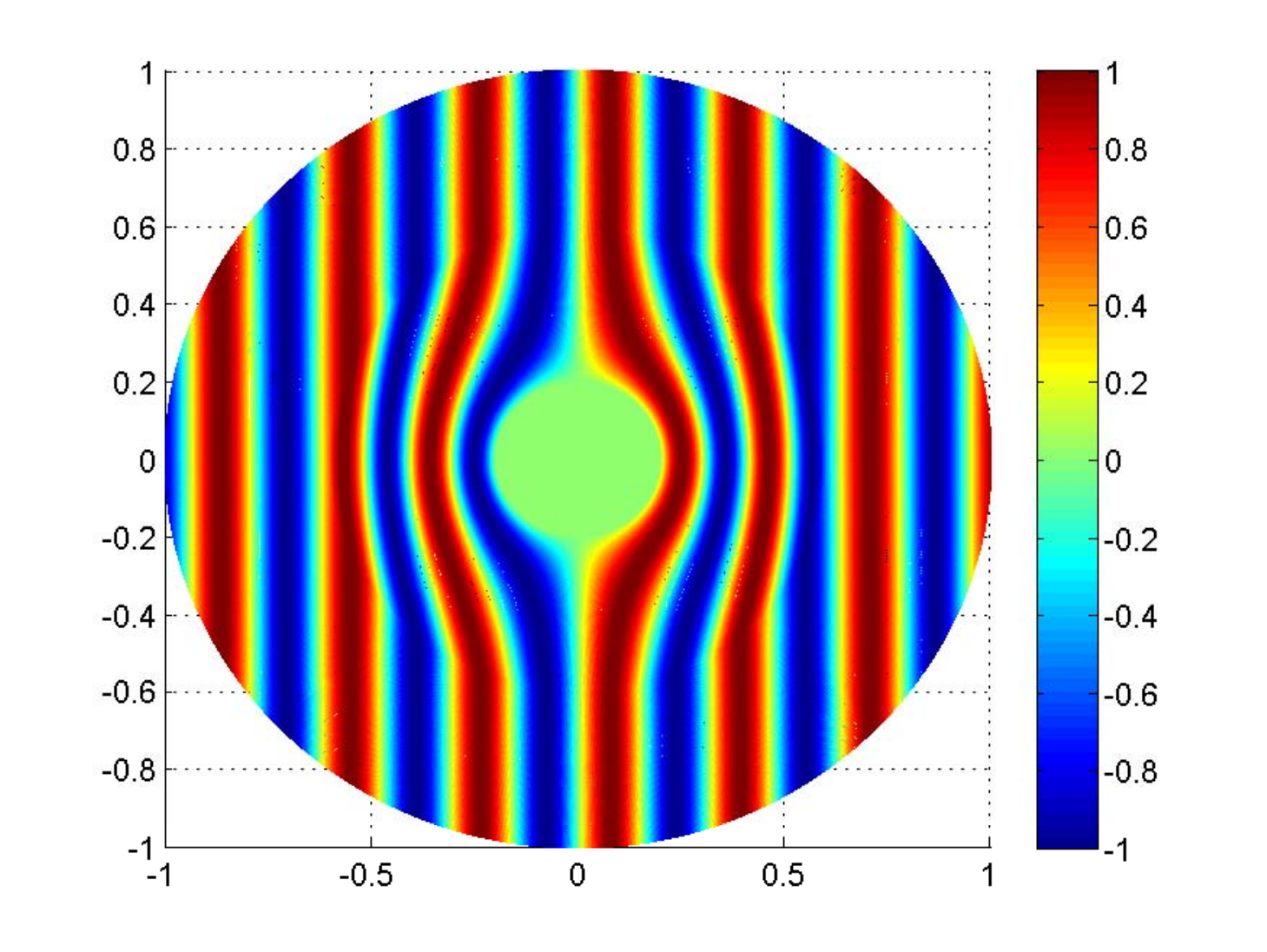}
   \includegraphics[width=0.87\textwidth,height=0.3\textwidth]{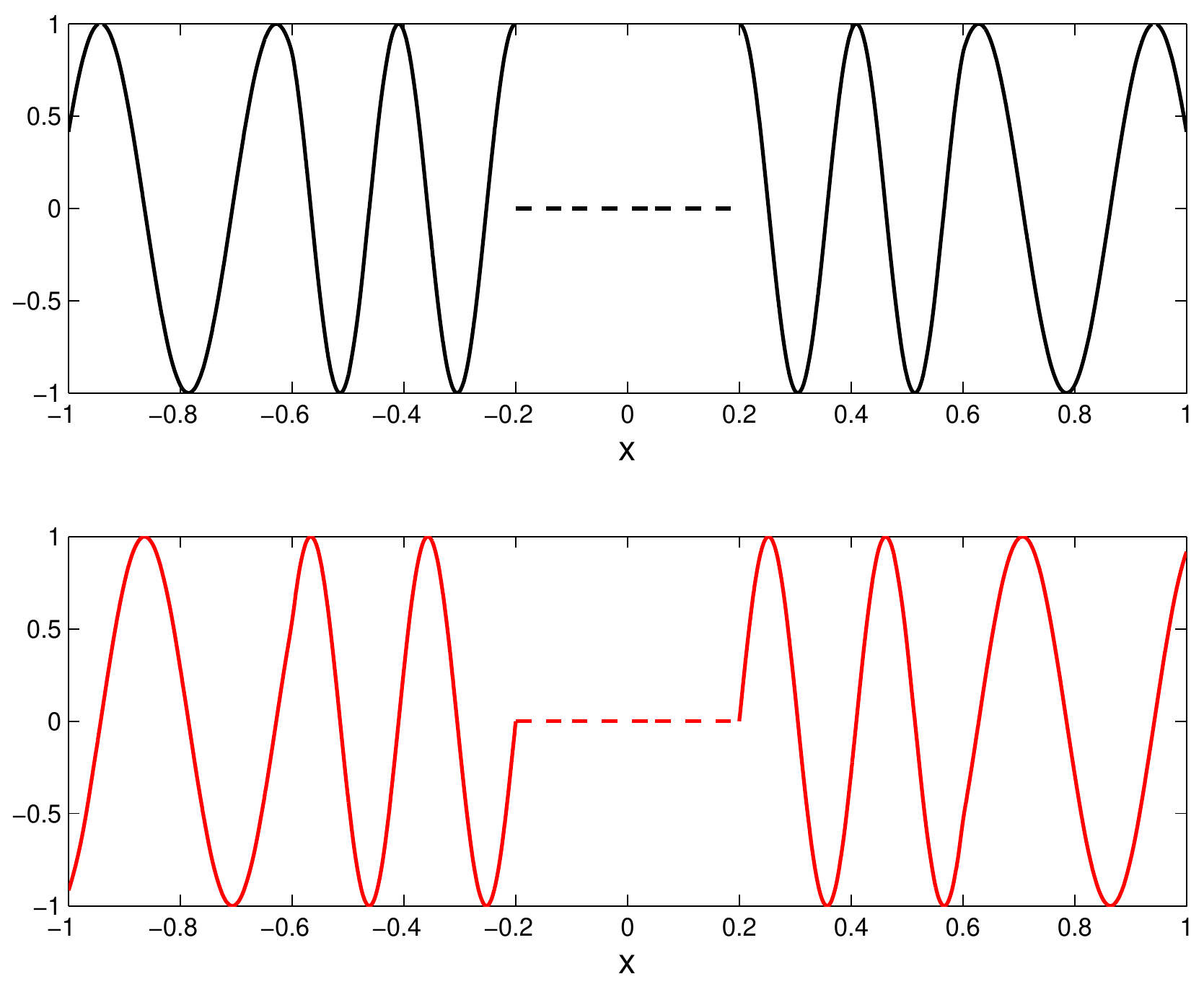}
   \caption{Row 1:  real (left) and imaginary (right) parts of the electric-field distributions.
   Row 2-3:  profiles of  the real  and imaginary  parts of the electric-field along $x$-axis. }
    \label{cloakk20}
\end{figure}

To further demonstrate the  performance of the proposed approach, we set  the incident angle
$\theta_0=\pi/3,$  increase the incident frequency to $k=100$ and enlarge  the cloaked region by taking
 $(R_1, R_2, R_3)=(0.3, 0.9, 1.0).$  We depict in Figure \ref{cloakk100}  the same type of numerical results  (obtained by the FLSEM with $M=120$ and $\bs N=(100,20)$) as in
    Figure \ref{cloakk20}.  Again, the highly oscillatory oblique incident wave is perfectly steered by the cloaking layer,  and  completely  shielded from the cloaked region.   It is also worthwhile to point out that  the exact boundary condition can be placed as close as possible to the cloak that can significantly reduce the number of grid points in the outermost artificial  shell, especially when the incident frequency is high.
%
%

\subsection{Wave generated by an external source}\label{external}
We now use  an external source,  compactly supported in the annulus $R_2<r<R_3,$ as  the wavemaker, and turn off the incident wave. More precisely, we modify   \eqref{domain2}  as
\begin{equation}
{\mathcal L}_0[u^2]=f  \;\;\;  {\rm in}\;\;  \Omega_2;\quad (\partial_r -{\mathcal T} _{R_3})u^2 =0   \;\;\; {\rm at}\;\;  r=R_3. \label{Sourceformu}
\end{equation}
In this situation, there is no  closed-form exact solution.
In practice, we use the Guassian function in Cartesian coordinates:
\begin{equation}\label{source_example1}
f(x,y)=\alpha\; {\rm exp}\Big({-\frac{(x-\beta)^2+(y-\kappa)^2}{2 \gamma^2}}\Big),
\end{equation}
where $\alpha,\beta,\kappa,\gamma$ are tuneable  constants, and $\gamma$ is small.
In the computation, we take  $(R_1, R_2, R_3)=(0.2, 0.6, 1.0)$,  $\alpha=100$, $\beta=-0.8$, $\kappa=0$ and $\gamma=0.02.$ The source at $r=R_3$ is nearly zero.  The plots of the electric-field distributions in Figure \ref{source} are computed from the FLSEM  with $k=40$, $M=40$ and $\bs N=(40,150).$   The non-plane  waves generated by the source are smoothly bent and the cloak does not produce any scattering.  We also observe from Figure \ref{source}  that the waves seamlessly pass through the outer artificial boundary without any reflecting.

\begin{figure}[h!]
  \centering
    \includegraphics[width=0.49\textwidth]{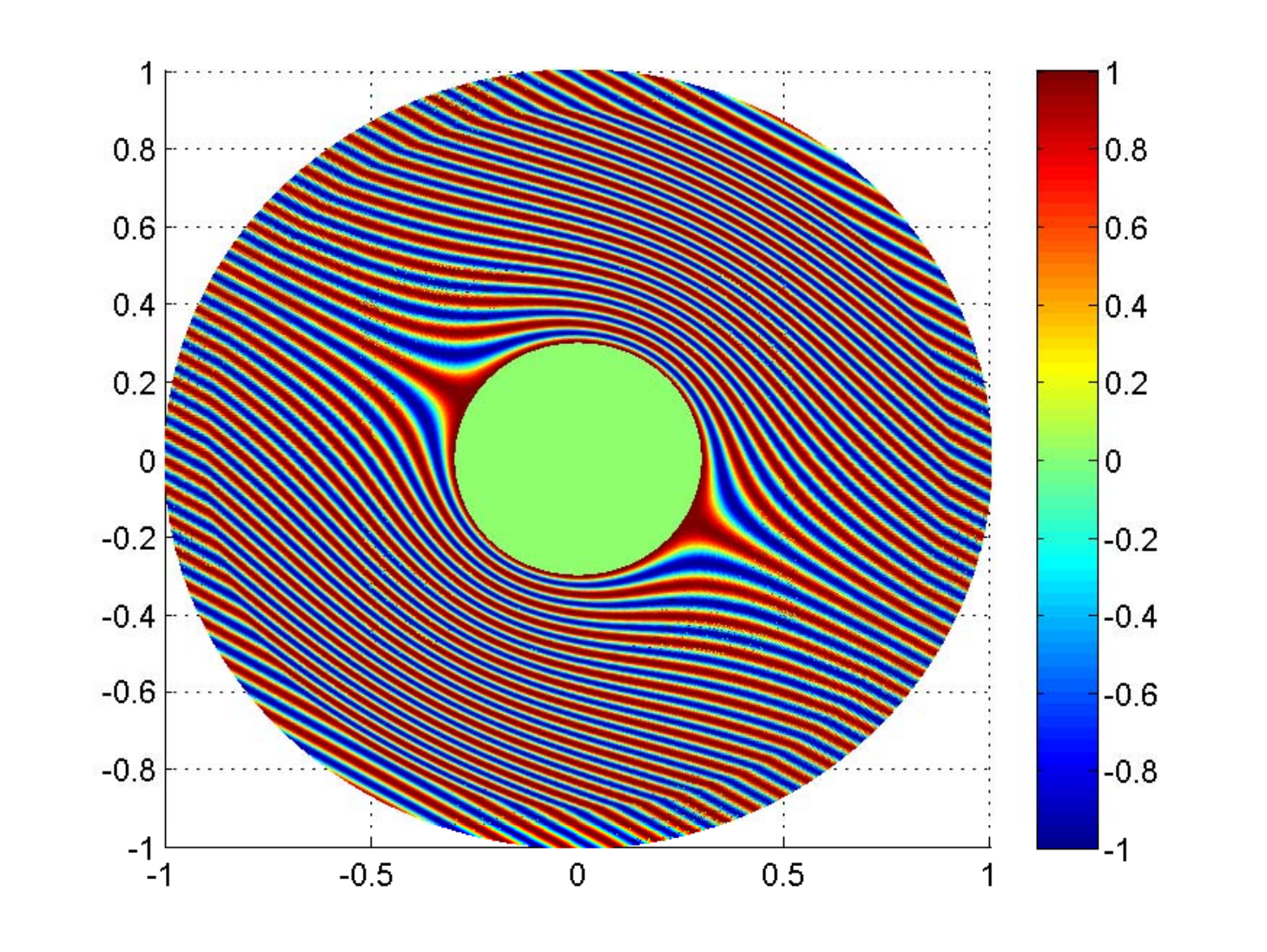}\hspace*{-18pt}
    \includegraphics[width=0.49\textwidth]{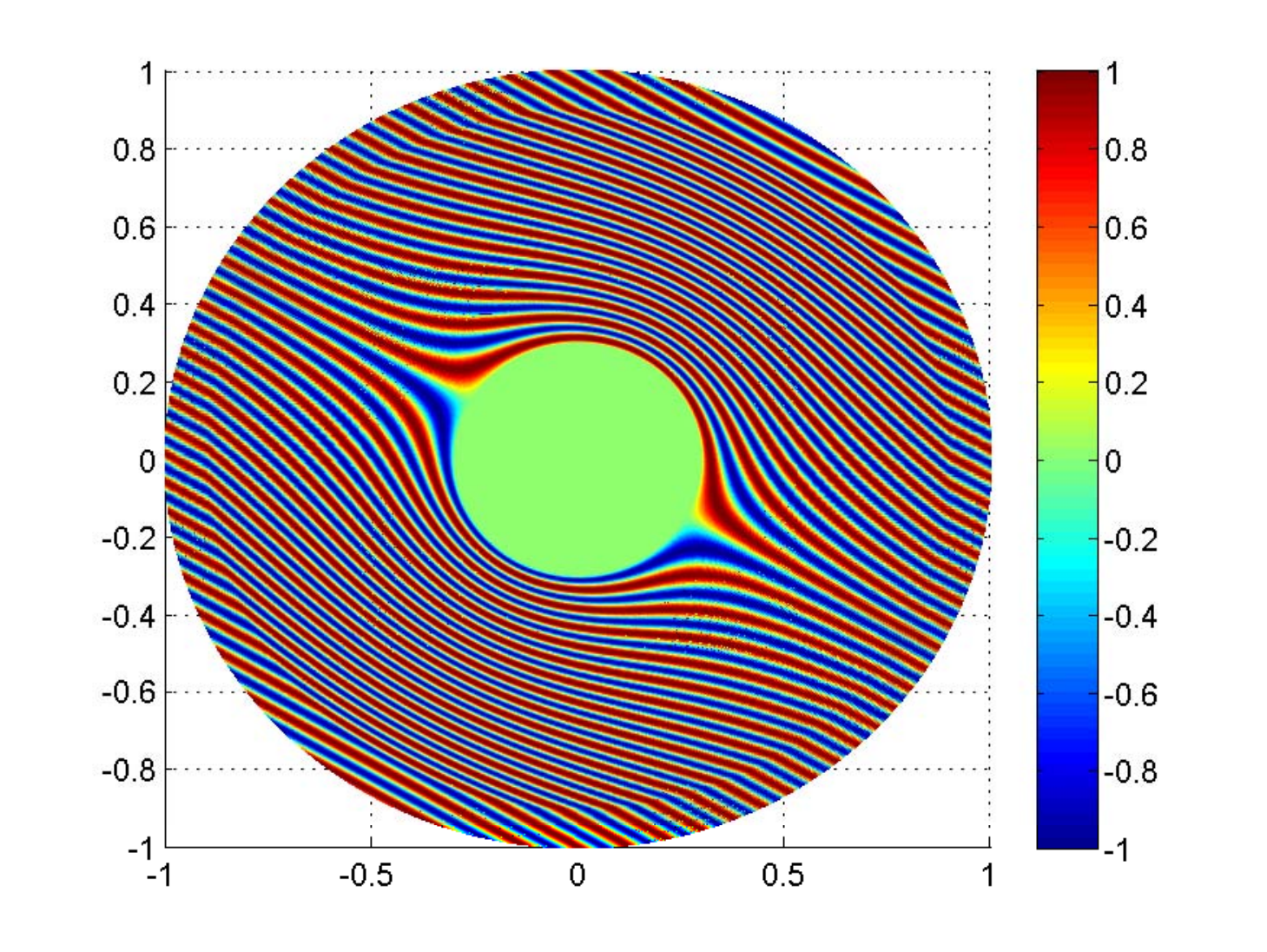}
 \includegraphics[width=0.87\textwidth,height=0.3\textwidth]{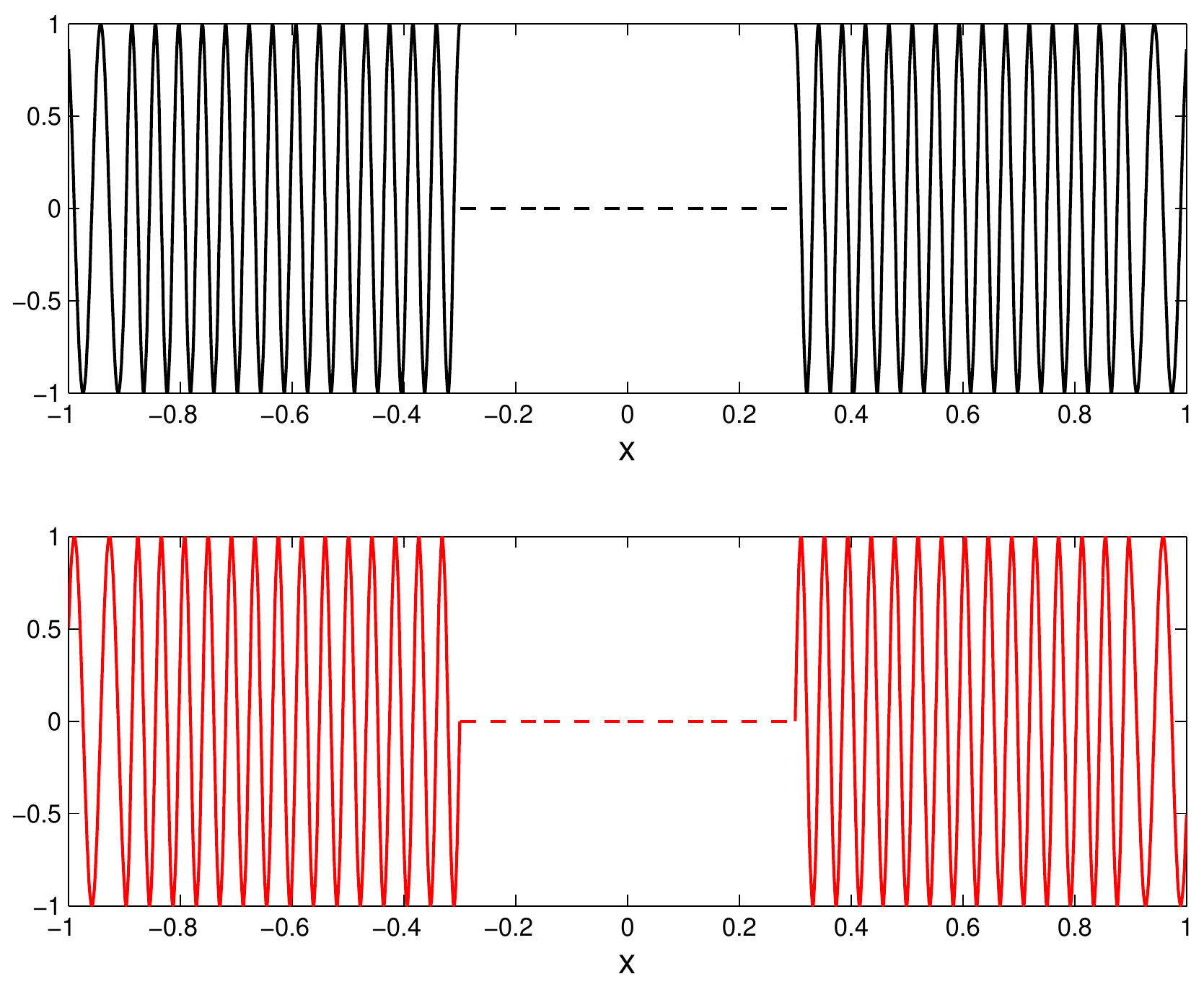}
   \caption{Row 1:  real (left) and imaginary (right) parts of the electric-field distributions.
   Row 2-3:  profiles of  the real  and imaginary  parts of the electric-field along  $\theta=\pi/3$.
   }
    \label{cloakk100}
\end{figure}

\begin{figure}[h!]
  \centering
    \includegraphics[width=0.49\textwidth]{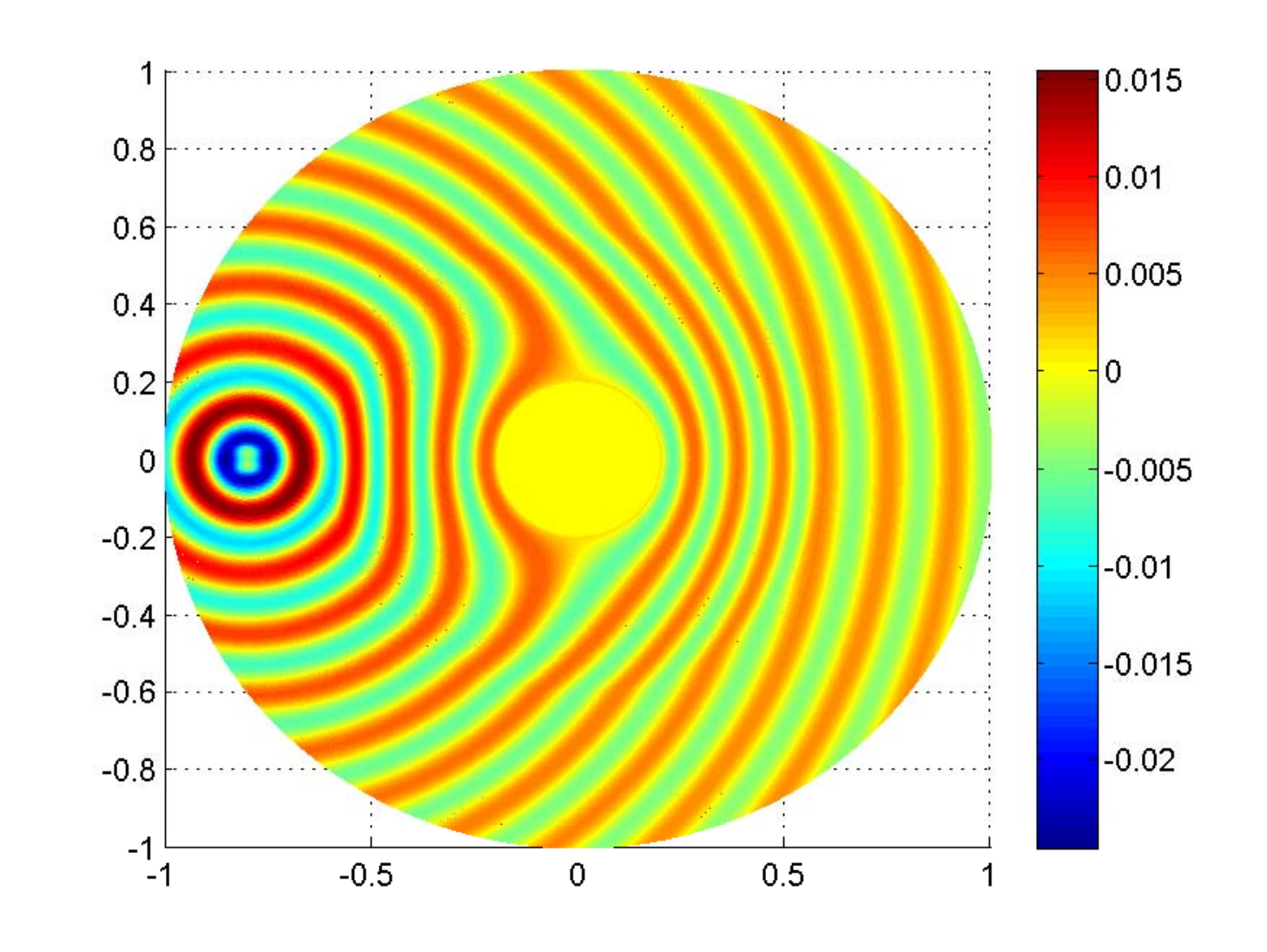}\hspace*{-15pt}
     \includegraphics[width=0.49\textwidth]{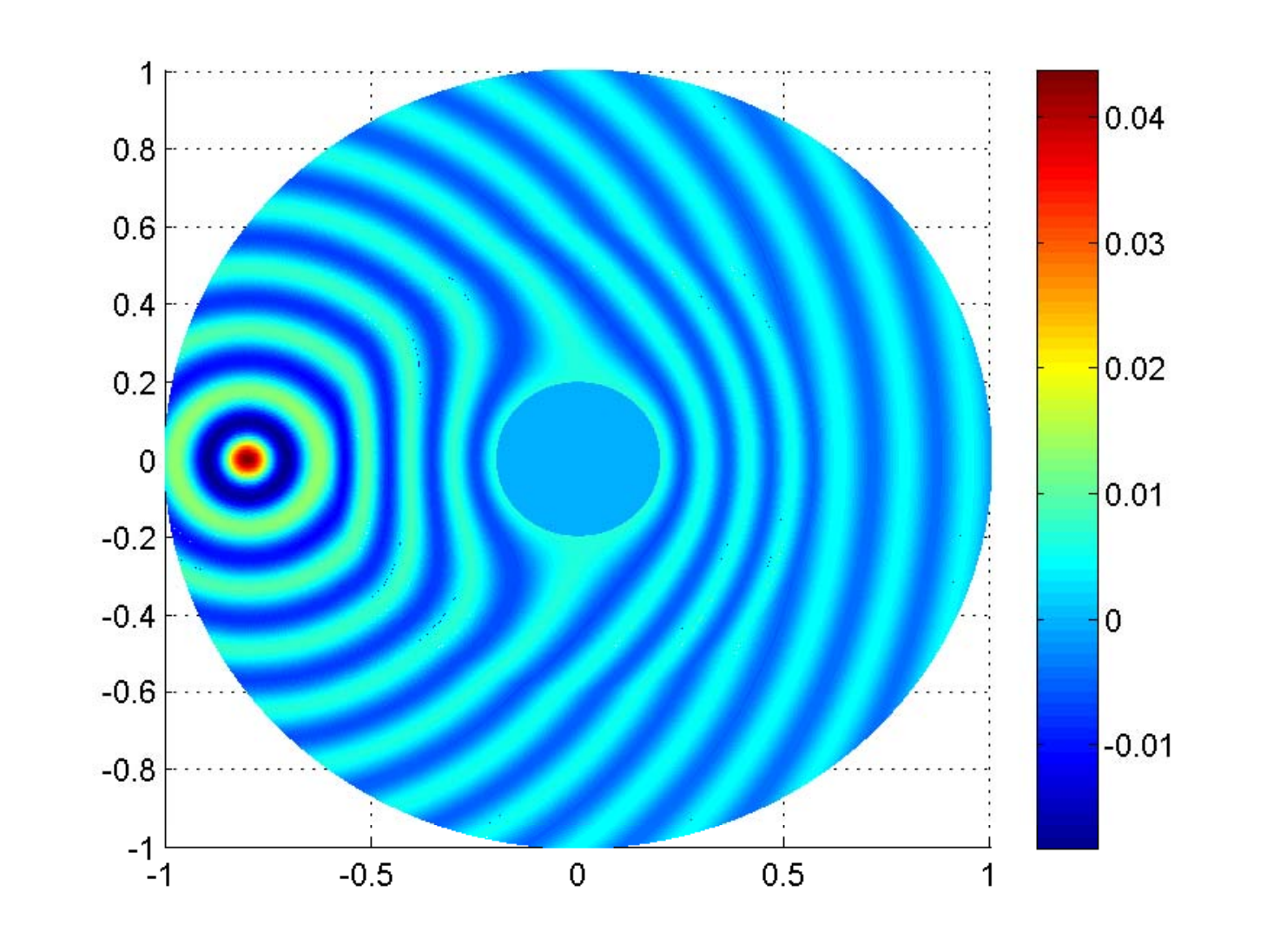}\hspace*{-15pt}
   \caption{Electric-field distributions with an external source compactly supported in the outmost shell.  Left: real part;  right: imaginary part. }
    \label{source}
\end{figure}

\subsection{Elliptic cyindrical  cloaks}
We first consider an incident plane wave in   \eqref{uinplane}, which, in the elliptic coordinates  (cf. \eqref{ellpco}),  can be expanded in terms of Mathieu functions
(cf. \cite[P. 218]{mechel2002formulas}):
\begin{equation}\label{ellipwave}
\begin{split}
v_{\rm in}(\xi,\eta)&={\rm exp}({{\rm i}ka(\cosh\xi \cos \eta \cos \theta_0+\sinh \xi \sin \eta \sin \theta_0)})\\
&= \sqrt{\frac{8}{\pi}}\sum_{m=0}^{\infty} \ri^{m} {\rm Mc}_m^{(1)}(\xi;q){\rm ce}_m(\theta_0;q) {\rm ce}_m(\eta;q)\\
                     &\quad +\sqrt{\frac{8}{\pi}}\sum_{m=1}^{\infty} \ri^{m} {\rm Ms}_m^{(1)}(\xi;q){\rm se}_m(\theta_0;q) {\rm se}_m(\eta;q).
\end{split}
\end{equation}
Note that when $q=0,$ it reduces to \eqref{uinplane}. Following  \cite{cojocaru2009exact}, we obtain  the exact solution for the ideal elliptic cloak similar to the circular case in   \eqref{uexactA}:
\begin{equation}\label{uexactAB}
v(\xi,\eta)
=\begin{cases}
v_{{\rm in}}(d(\xi-\xi_1),\eta) ,\quad & {\rm if}\;\;  \xi_1< \xi <\xi_2,\\
v_{{\rm in}}(\xi,\eta) ,\quad & {\rm if}\;\;  \xi> \xi_2,
\end{cases}
\end{equation}
which vanishes if $0<\xi<\xi_1.$

To illustrate the spectral accuracy of  MLSEM, we tabulate in Table \ref{table1}
the numerical errors $E_{\bs N}$ (defined as in  the circular case) for different $\bs N=(N,N)$ and for several $k$. In the simulation,  we take $a=0.6, \theta=0,  (\xi_1, \xi_2, \xi_3)=(0.7,1.3,1.5)$ and $M=70.$

\begin{table}[!h]
\caption{Convergence of MLSEM}
\begin{center}
\setlength{\tabcolsep}{20pt}
\begin{tabular}{llll}
\hline
       & $k=30$      & $k=50$      &   $k=70$ \\
\cline{2-4}
  $N$          & error      & error      &   error  \\
\hline
 30         & 8.63E-07    & 2.75E-02    &   1.32   \\
 40         & 4.66E-12    & 8.42E-06    &   8.50E-02  \\
 50         & 9.03E-15    & 4.87E-10    &   3.52E-05 \\
 60         & 1.57E-14    & 2.27E-14    &   5.89E-09  \\
 70         & 1.66E-14    & 2.08E-14    &   2.73E-13  \\
\hline
\end{tabular}
\end{center}\label{table1}
\end{table}

We next  illustrate    electric-field distributions. The circular cloak is perfectly symmetric, so the way of bending the waves is independent of
the incident angle.  However, as pointed out in,  e.g.,  \cite{ma2008material},   the incident wave along the major axis (i.e., $\theta_0=0$) leads to significantly better cloaking effect with much less scattering and bears the greatest resemblance  to the circular cloak,  compared with  other directions.
Note that in  \cite{ma2008material}, PEC condition  was imposed at the inner boundary $\xi=\xi_3$ in the finite-element  simulation, and the electric-field distributions exhibited  observable scattering waves when the incident angle $\theta_0\not=0.$  However, using our proposed CBCs and numerical solver,  the perfect cloaking effect can be achieved equally and no any scattering is induced for any incident angles.  Apart from plotting
the electric-field distributions, we  also depict the  time-averaged Poynting vector (cf. \cite{orfanidis2002electromagnetic}):
      \begin{equation}\label{poynting}
{\bs S}={\rm Re}\,\{{\bs E}\times {\bs H}^*\}/2,
\end{equation}
which indicates the directional  energy flux density.  In Figures \ref{Elliptheta0}-\ref{Ellipthetap}
(where $\theta_0=0, \pi/4, \pi$, respectively, and in all cases, $a=0.6$, $(\xi_1, \xi_2, \xi_3)=(0.7, 1.3, 1.5), k=20$, $M=30$ and $\bs N=(20,20)$),  we plot the real part of the electric-field distributions (note: the imaginary part behaves very similarly), and the corresponding Poynting vector fields.
 We find that the waves are again  steered smoothly around the elliptic cloaked region without reflecting and scattering.  We particularly look at the Poynting vectors in Figure \ref{Ellipsource}, where the energy flux attempts to flow  across $r=\xi_1,$  but it is directed by the cloak. Once again,  the surface current is  induced on the cloaking interface as with the circular cloak.   It is noteworthy  that   the incident wave perpendicular to the major axis (see Figure \ref{Ellipsource})  is of
particular interest,   as the shape is like a slap and the waves are difficult to steer  (cf. \cite{jiang2008arbitrarily}).
However, using our approach, the perfect concealment of waves can be achieved as with other incident angles.

Finally, we  conduct a  test  by adding an external source (cf. Subsection \ref{external}), and turning  off the incident wave.  Accordingly, we modify \eqref{2DHelm_Ellip2} and \eqref{ellipticexact} as
\begin{align}\label{externalforce}
&\big(\partial^2_{\xi} +\partial ^2 _{\eta}\big)v +k^2 a^2 \big(\cosh^2 \xi-\cos^2 \eta \big)v=f \;\;   {\rm in}\;\;    \Lambda_2;\;\;   (\partial_{\xi} -{\mathbb T} _{\xi_3}) v=0  \;\;   {\rm at}\;\;  \xi=\xi_3,
\end{align}
where  $f$  is compactly supported in the elliptic layer $\xi_2<\xi<\xi_3$. Like before, we take $f$ to be \eqref{source_example1} with $\alpha=1000$, $\beta=0$, $\kappa=1.148$, $\gamma=0.01$.
Figure \ref{Ellipsource} is computed from MLSEM with $k=20$, $M=30$ and $\bs N=(50,100)$ and illustrates the real (left) and imaginary (right) part of the electric-field distributions induced by external source.
Once again, we see that the waves  are smoothly bent  without penetrating into the elliptic cloaked region.
Moreover, the cloak does not induce any scattering, and  indeed, we see   the fields near the source  totally unaffected.



\begin{figure}[h!]
\begin{minipage}{0.6\textwidth}
  \centering
    \includegraphics[width=1\textwidth]{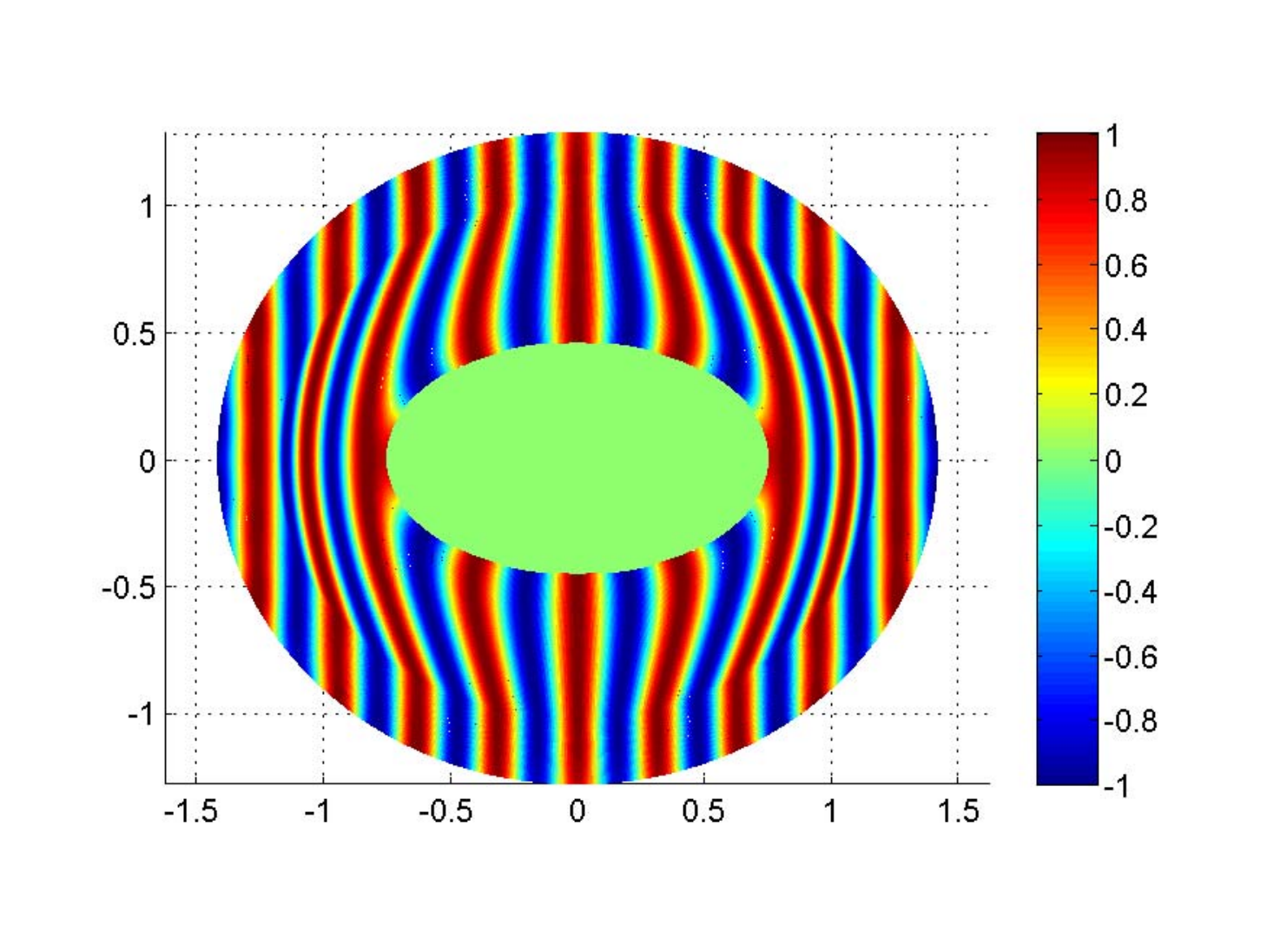}
  \end{minipage}
  \hspace*{-16pt}
\begin{minipage}{0.40\textwidth}
  \centering
     \includegraphics[width=0.88\textwidth]{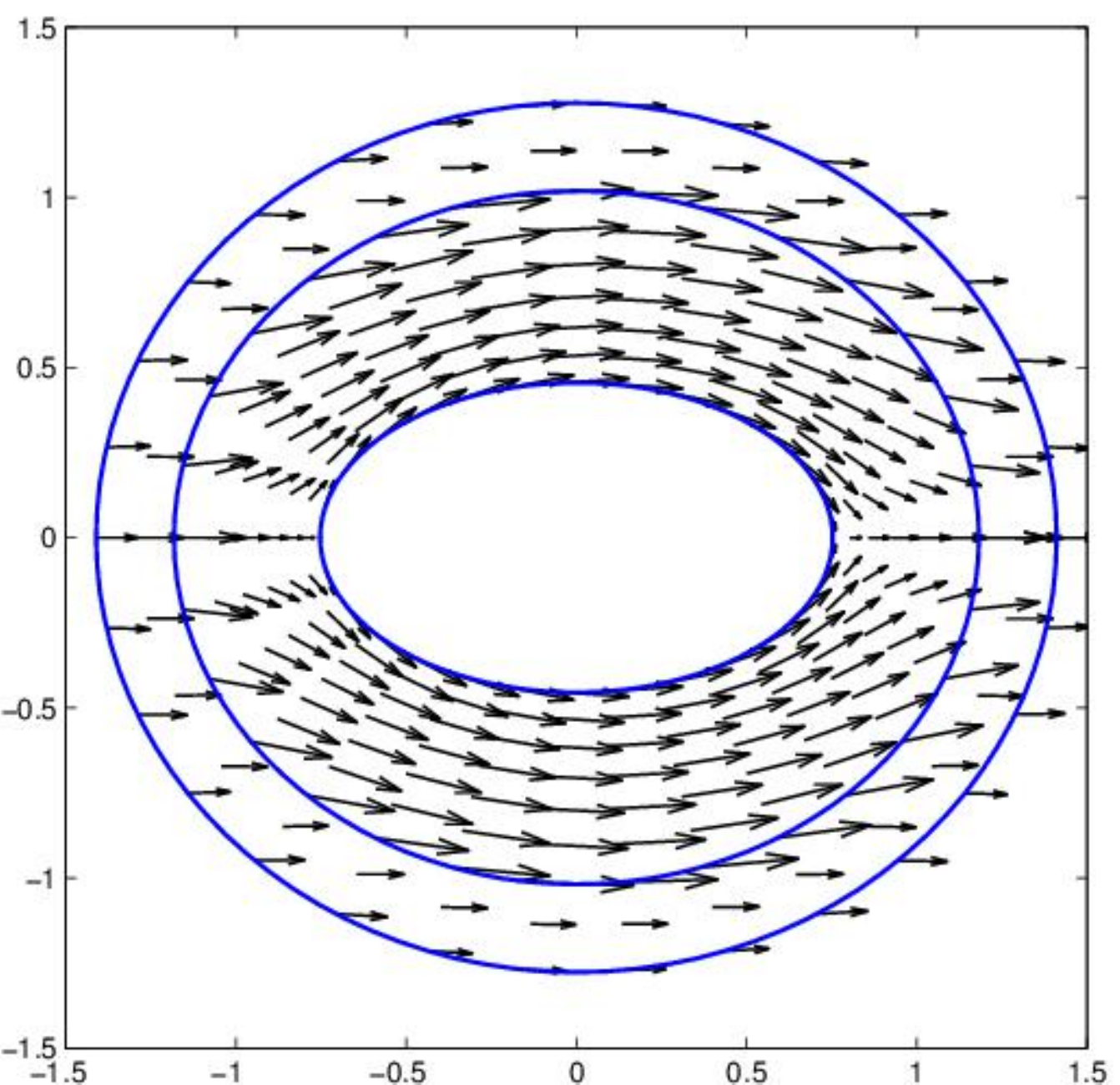}
  \end{minipage}
  \vspace*{-15pt}
   \caption{Real part of the  electric-field distribution (left) and the related Poynting vector (right), where the incident angle  $\theta_0=0$.}
    \label{Elliptheta0}
\end{figure}

\vspace*{-20pt}
\begin{figure}[h!]
\begin{minipage}{0.6\textwidth}
  \centering
    \includegraphics[width=1\textwidth]{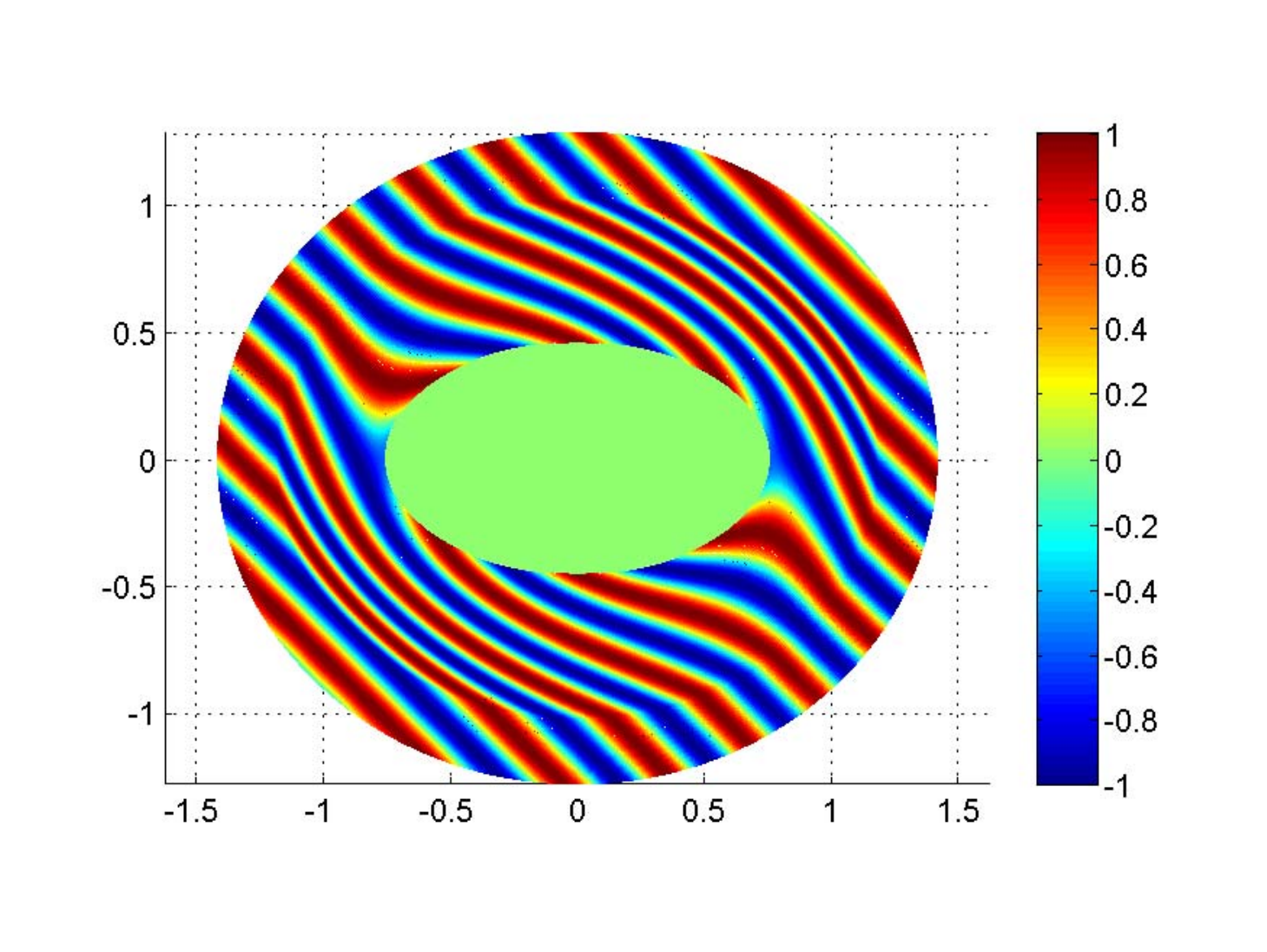}
  \end{minipage}
  \hspace*{-16pt}
\begin{minipage}{0.4\textwidth}
  \centering
     \includegraphics[width=0.88\textwidth]{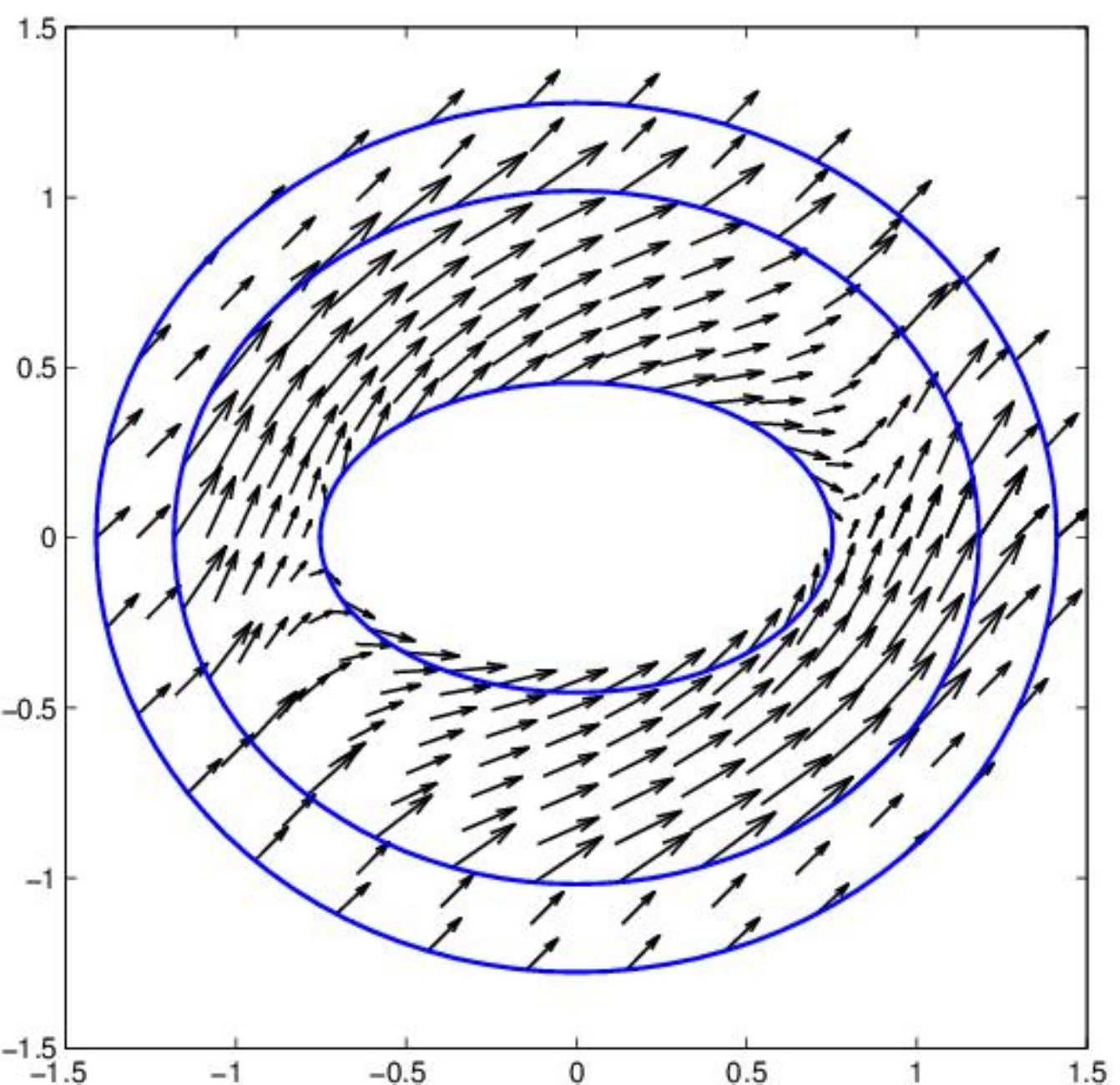}
  \end{minipage}
  \vspace*{-15pt}
   \caption{Real part of the  electric-field distribution (left) and the related Poynting vector (right), where the incident angle  $\theta_0={\pi}/{4}$. }
    \label{Ellipthetapi4}
\end{figure}

\vspace*{-20pt}

\begin{figure}[h!]
\begin{minipage}{0.6\textwidth}
  \centering
    \includegraphics[width=1\textwidth]{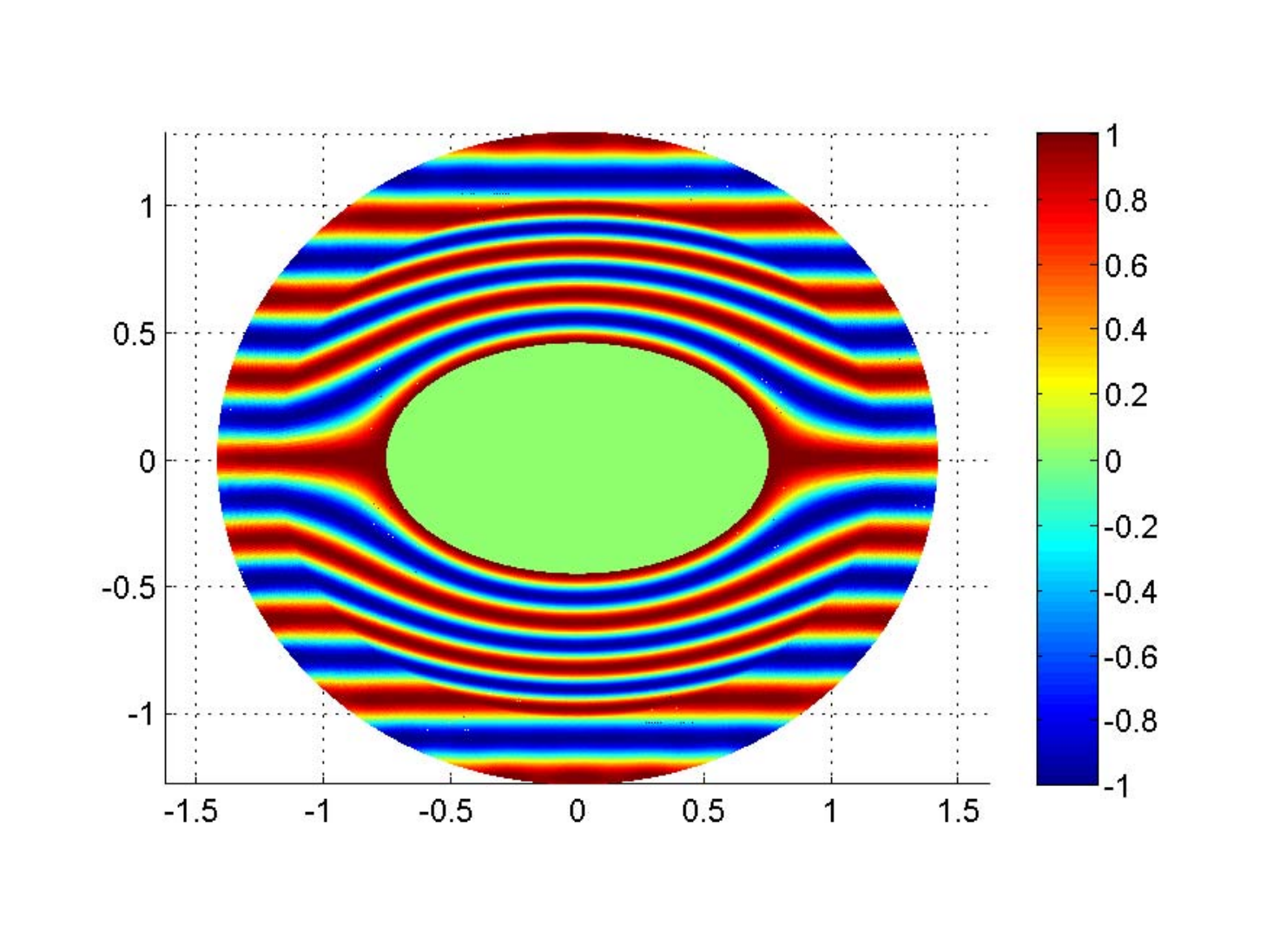}
  \end{minipage}
  \hspace*{-16pt}
\begin{minipage}{0.4\textwidth}
  \centering
     \includegraphics[width=0.88\textwidth]{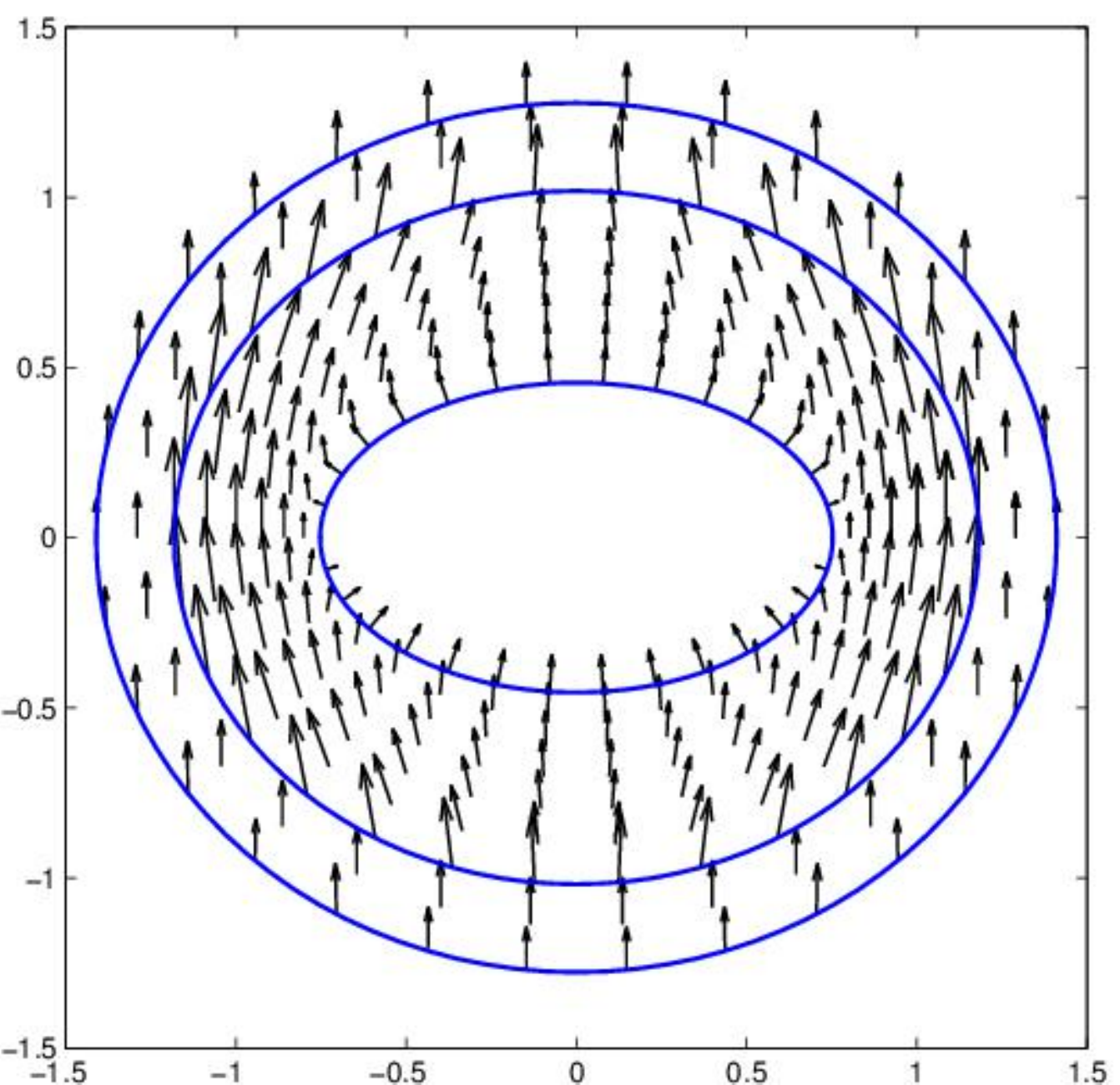}
  \end{minipage}
  \vspace*{-15pt}
   \caption{Real part of the  electric-field distribution (left) and the related Poynting vector (right), where the incident angle  $\theta_0={\pi}$. }
    \label{Ellipthetap}
\end{figure}

\begin{figure}[h!]
  \centering
    \includegraphics[width=0.51\textwidth]{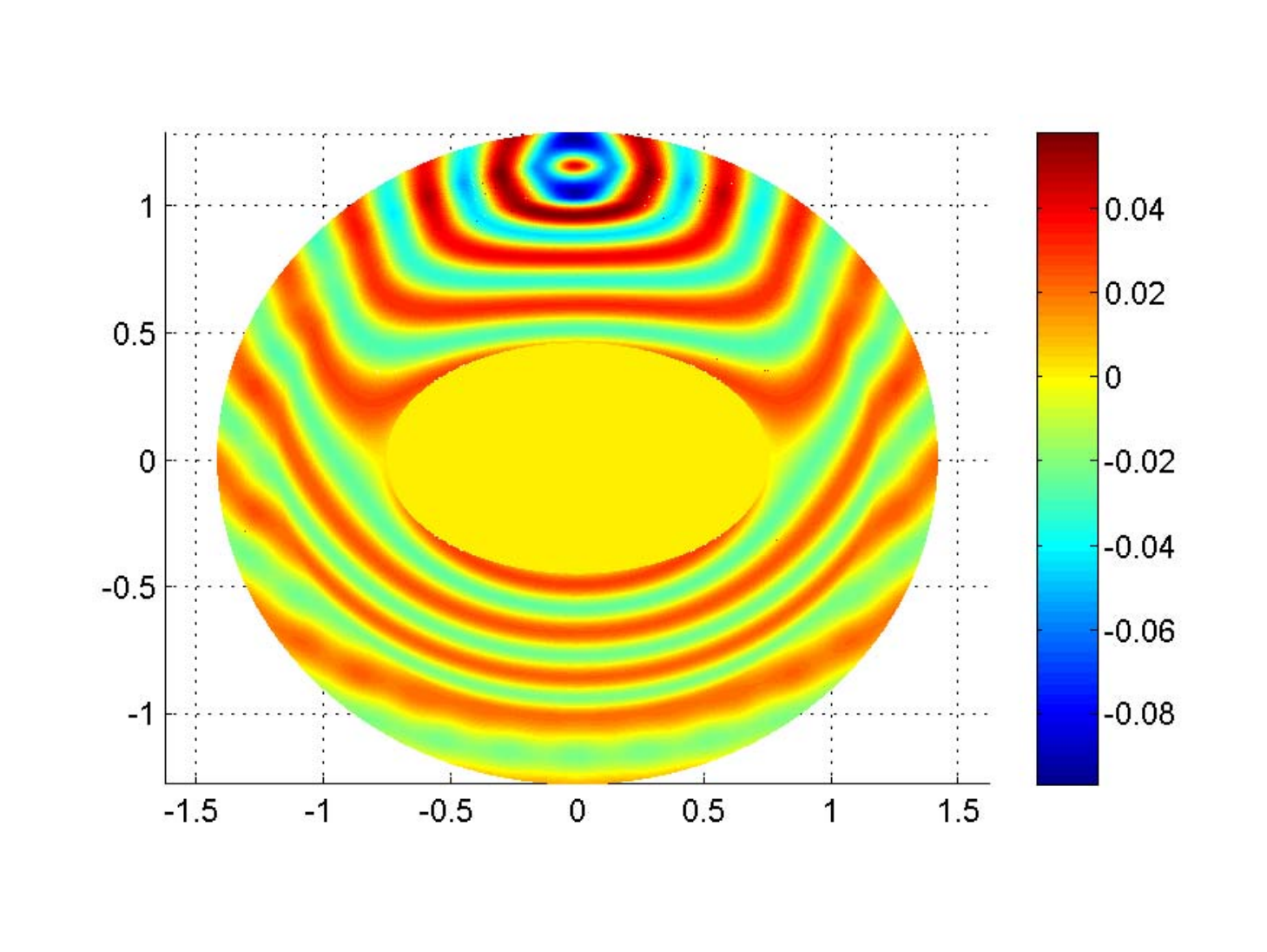}
    \hspace*{-18pt}
    \includegraphics[width=0.51\textwidth]{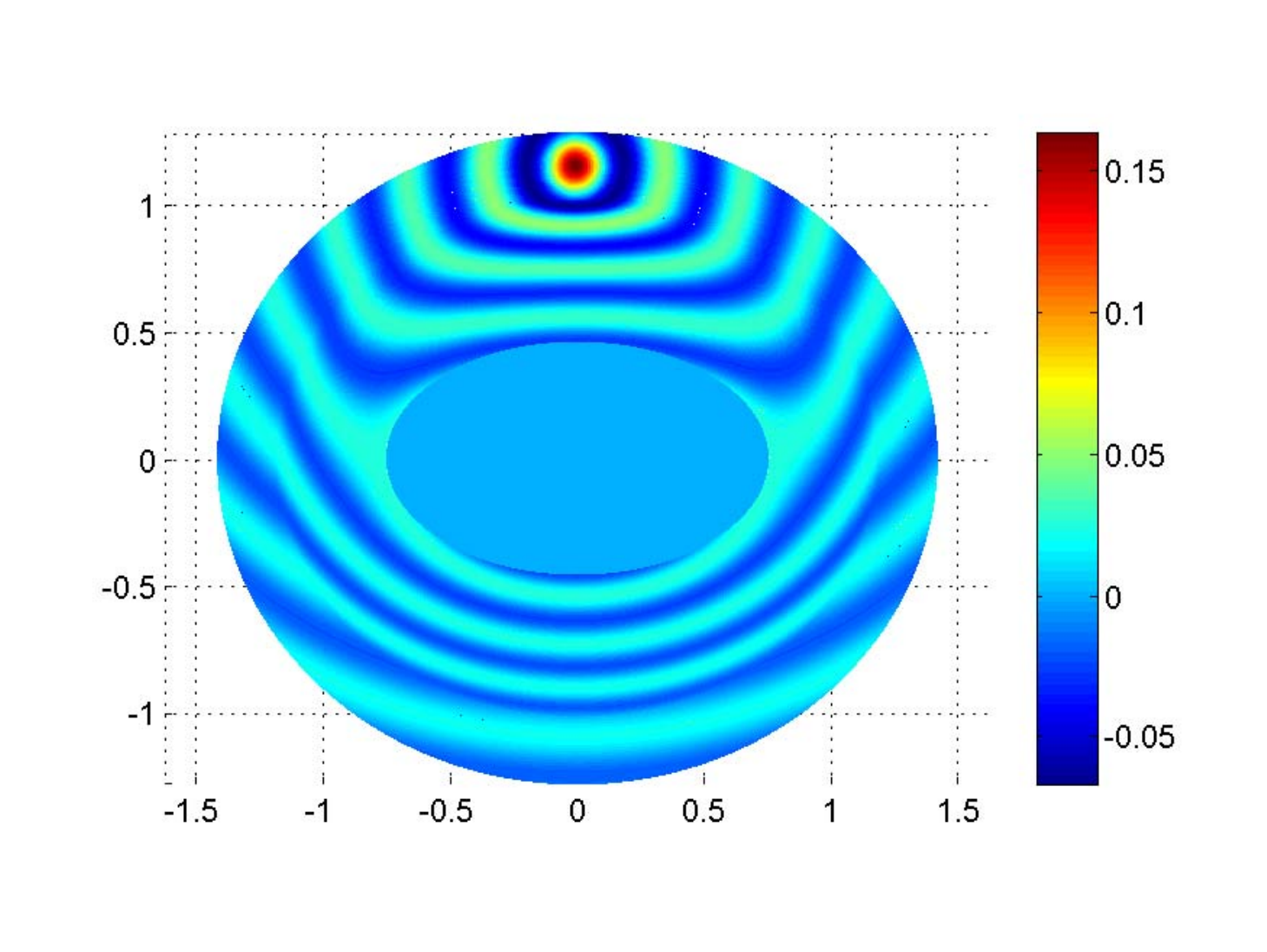}
   \caption{The real (left) and imaginary (right) part of electric-field distribution with different sources.}
    \label{Ellipsource}
\end{figure}

\vskip 40pt

\vskip 10pt
\noindent\underline{\large\bf Concluding remarks}
\vskip 10pt

From a new perspective,  we proposed CBCs for the ideal circular and elliptic cylindrical cloaks, which, together with a super-accurate spectral-element solver,  demonstrated that the cloaks can achieve perfect concealment of
incoming incident waves with very mild conditions on the incident frequency. We also illustrated the perfect cloaking effect, when the incoming wave is generated by a source exterior to the cloaking device.

The idea and approach in this paper  can shed light on the study of, e.g., polygonal cloaks \cite{Diatta2009,Wu2009material, Luo2011arbitrary}, and can lead to appropriate CBCs for time-domain simulations  \cite{Hao2008fdtd,li2012time}.

\vskip 10pt
\noindent\underline{\large\bf Acknowledgment}
\vskip 6pt
The authors would like to thank Professor Jichun Li from University of Nevada, Las Vegas, USA, and  Professors Baile Zhang and Handong Sun in the Division of Physics and Applied Physics of Nanyang Technological University, for fruitful discussions.

\appendix
\renewcommand{\thesection}{\Alph{section}}
\renewcommand{\theequation}{\thesection.\arabic{equation}}
\section{Proof of Proposition $2.2$}\label{prop:uniquesolu}
  It is clear that by    \eqref{Iweakform},
\begin{equation}\label{realpart}
{\rm Re}\big\{{\mathcal B}_m(\hat u_m, \hat u_m)\big\}=\|\hat u_m'\|_\omega^2+m^2\|\hat u_m\|_{\omega^{-1}}^2-k^2\|\rho\hat u_m\|^2_{\omega} - R_3\, {\rm Re}(\mathcal T_{m,k})|\hat u_m(R_3)|^2,
\end{equation}
and
\begin{equation}\label{imagpart}
{\rm Im}\big\{{\mathcal B}_m(\hat u_m, \hat u_m)\big\}= -R_3\, {\rm Im}(\mathcal T_{m,k}) |\hat u_m(R_3)|^2.
\end{equation}
Recall that  (see, e.g.,   \cite{Nede01,She.W07}):
\begin{equation}\label{kernelprop}
{\rm Re} (\mathcal T_{m,k})<0,\quad  {\rm Im} (\mathcal T_{m,k})>0.
\end{equation}
Thus, we have
\begin{equation}\label{realpart2A}
{\rm Re}\big\{{\mathcal B}_m(\hat u_m, \hat u_m)\big\}\ge \|\hat u_m'\|_\omega^2+m^2\|\hat u_m\|_{\omega^{-1}}^2-k^2\|\rho\hat u_m\|^2_{\omega},
\end{equation}
and $\hat u_m(R_3)=0,$ if $\hat g_m=0.$
Using  the Fredholm alternative (cf.  \cite[Thm. 5.4.5]{Nede01}),  we reach the conclusion.

\renewcommand{\thesection}{\Alph{section}}
\renewcommand{\theequation}{\thesection.\arabic{equation}}
\section{Proof of Proposition $3.2$}\label{AppendixA}

  For simplicity,  denote
\begin{equation*}
{\rm M}_m^{(i)}={\rm Mc}_m^{(i)}\;\; {\rm or}\;\; {\rm Ms}_m^{(i)},\quad i=1,2,3;\quad  {\mathcal D} _{m}={\mathcal D} _{m}^{\rm c}\;\; {\rm or}\;\; {\mathcal D} _{m}^{\rm s}.
\end{equation*}
Recall that  (see e.g., \cite{Abr.S84})
\begin{equation}\label{properMathieu}
{\rm M}_m^{(3)}:={\rm M}_m^{(1)}+\ri {\rm M}_m^{(2)}; \quad {\rm M}_m^{(1)} {{\rm M}_m^{(2)}}'  - {\rm M}_m^{(2)} {{\rm M}_m^{(1)}}'=\frac{2}{\pi}\,.
\end{equation}
Then a direct calculation from  \eqref{senewcond2A} and \eqref{properMathieu} leads to
\begin{equation}\label{Dmimag}
{\rm Im}({\mathcal D} _{m})=\frac{{\rm M}_m^{(1)}(\xi_3; q){{\rm M}_m^{(2)}}'(\xi_3; q)-{\rm M}_m^{(2)}(\xi_3; q){{\rm M}_m^{(1)}}'(\xi_3; q)}{|{\rm M}_m^{(3)}(\xi_3; q)|^2}=\frac{2/\pi}{ |{\rm M}_m^{(3)}(\xi_3; q)|^2}>0.
\end{equation}
Moreover, by \eqref{1DceD},
\begin{equation}\label{Dmreal}
{\rm Re}({\mathcal D} _{m})=\frac{{\rm M}_m^{(1)}(\xi_3; q){{\rm M}_m^{(1)}}'(\xi_3; q)+{\rm M}_m^{(2)}(\xi_3; q){{\rm M}_m^{(2)}}'(\xi_3; q)} {|{\rm M}_m^{(3)}(\xi_3; q)|^2},
\end{equation}
Note that  $\{{\rm M}_m^{(i)}(\xi;q)\}_{i=1}^2$ can not have common zero,  and are analytic for all $\xi>0$ (see,  e.g., \cite{Abr.S84}), so   $|{\rm Re}({\mathcal D} _{m})|$ is a finite constant for fixed $m,q$ and $\xi_3.$ To this end, let $C$ be a generic positive constant depending on $m,q, \xi_2$ and $\xi_3.$

We first consider \eqref{EllipWeakc} and obtain that
\begin{equation}\label{Erealpart}
{\rm Re}\big\{{\mathcal B}_m^{\rm c}(\hat v_m^{\rm c}, \hat v_m^{\rm c})\big\}=  \| \varpi(  \hat v_m^{\rm c})' \|^2+ \tilde \lambda_m^{\rm c} \| \varpi^{-1} \hat v_m^{\rm c}\|^2-4q \|\varpi^{-1}{\chi} \hat v_m^{\rm c} \|^2
-{\rm Re}({\mathcal D}_m^{\rm c})|\hat v_m^{\rm c}(\xi_3)  |^2,
\end{equation}
and
\begin{equation}\label{Eimagpart}
{\rm Im}\big\{{\mathcal B}_m^{\rm c}(\hat v_m^{\rm c}, \hat v_m^{\rm c})\big\}= -{\rm Im}({\mathcal D}_m^{\rm c})|\hat v_m^{\rm c}(\xi_3)  |^2.
\end{equation}
Let $I_2=(\xi_2,\xi_3)$ as before.  Recall the Sobolev inequality (see,  e.g.,  \cite[(B.33)]{ShenTaoWang2011}): for any $w\in H^1(I_2),$
\begin{equation}\label{soblevineq}
\max_{x\in \bar I_2}|w(x)|^2\le \Big(\frac 1 {\xi_3-\xi_2}+2\Big)\|w\|_{L^2(I_2)}\|w\|_{H^1(I_2)}.
\end{equation}
Therefore,  we further derive from the Cauchy-Schwartz inequality that
\begin{equation}\label{ImagIneq}
\begin{split}
|{\rm Re}({\mathcal D}_m^{\rm c})|&|\hat v_m^{\rm c}(\xi_3)  |^2 \leq C\big(\|\varpi (\hat v_m^{\rm c})'\|^2+\|\chi \hat v_m^{\rm c} \|^2\big)^{{1}/{2}}\|\chi \hat v_m^{\rm c}\|     \\
&\leq C\big(\|\varpi (\hat v_m^{\rm c})'\| \|\chi \hat v_m^{\rm c}\| +\|\chi \hat v_m^{\rm c} \|^2 \big)
 \leq \frac{1}{2}\|\varpi (\hat v_m^{\rm c})'\|^2+C\|\chi \hat v_m^{\rm c}\|^2.
\end{split}
\end{equation}
Therefore, by \eqref{Erealpart} and  \eqref{ImagIneq},  
\begin{equation}\label{realineq}
{\rm Re}\big\{{\mathcal B}_m^{\rm c}(\hat v_m^{\rm c}, \hat v_m^{\rm c})\big\} \ge \frac{1}{2}\| \varpi(  \hat v_m^{\rm c})' \|^2+ \tilde \lambda_m^{\rm c} \| \varpi^{-1} \hat v_m^{\rm c}\|^2-(4q+C) \|\varpi^{-1}{\chi} \hat v_m^{\rm c} \|^2.
\end{equation}
Moreover,  by \eqref{Dmimag},
 $\hat v_m^{\rm c}(\xi_3)=0,$ if $\hat \phi_m^{\rm c}=0.$
Using  the Fredholm alternative (see, e.g.,  \cite[Thm. 5.4.5]{Nede01}),  we reach the conclusion.

The uniqueness of the solution for \eqref{EllipWeaks} can be shown similarly.



\end{document}